\documentclass[letterpaper,twocolumn,10pt]{article}
\usepackage{usenix2019_v3}
\usepackage[available]{usenixbadges}

\usepackage{color}
\usepackage[dvipsnames]{xcolor}
\hypersetup{
    linkcolor=blue,   
    citecolor=red,    
    urlcolor=magenta  
}
\usepackage{algorithm}
\usepackage{appendix}
\usepackage{tikz}
\usepackage{amsmath}
\usepackage{amsthm}
\usepackage{listings}
\usepackage{bm}
\usepackage{xspace}
\usepackage{comment}
\usepackage{graphicx}
\usepackage{subcaption}
\usepackage{svg}
\usepackage{booktabs}
\usepackage[normalem]{ulem} 
\usepackage{sidecap}
\usepackage{xcolor}
\usepackage[font=small,labelfont=bf]{caption}
\usepackage{tcolorbox}
\usepackage{subcaption}

\usepackage[T1]{fontenc} 

\usepackage[noend]{algpseudocode}
\usepackage{amssymb}
\usepackage{multirow}

\usepackage{enumitem}
\setlist{topsep=0pt, nosep, leftmargin=*}

\DeclareGraphicsExtensions{.png,.pdf}

\newcommand{\sysname}{S4-FIFO\xspace}

\newcommand{\nope}[1]{}

\newcommand{\rev}[1]{{#1}}
\newcommand{\revv}[1]{{#1}}

\DeclareSymbolFont{cmletters}{OML}{cmm}{m}{it}
\DeclareMathSymbol{\tau}{\mathord}{cmletters}{28}
\DeclareMathSymbol{\rho}{\mathord}{cmletters}{26}
\DeclareMathSymbol{\kappa}{\mathord}{cmletters}{20}

\begin{document}
\pagestyle{empty}

\date{}

\title{\Large \bf Learning-Augmented Heuristics: Simple, yet Smart, Robust and Interpretable Cache Eviction}

\author{
{\rm Haocheng Xia}\\
Harvard University \& UIUC
\and
{\rm William Nixon}\\
University of Chicago \& Harvard University
\and
{\rm Bintang Dwi Marthen}\\
Harvard University \& Institut Teknologi Bandung 
\and
{\rm Pranav Bhandari}\\
Meta
\and
{\rm Juncheng Yang}\\
Harvard University
} 

\maketitle
\begin{abstract}
Caching is widely used across the system stack to improve performance and efficiency, with eviction algorithms at its core. 
Existing cache eviction policies fall into two broad categories: static heuristics (e.g., 2Q, S3-FIFO) and smart algorithms (e.g., ARC, LRB). Smart caches can adapt to workloads and have the potential to achieve higher efficiency and robustness than static heuristics. 
However, we find that existing smart caches suffer from objective mismatches and instability. 

We introduce Learning-Augmented Heuristics (LAH), a framework that learns the cache-level parameters of static heuristics. By decoupling the data and control planes, LAH supports simple, high-speed data reads and writes on the data plane, while performing occasional asynchronous learning on the control plane using cache-level features. 

We demonstrate the effectiveness of LAH through S4-FIFO, a \underline{S}mart \underline{S3}-FIFO cache eviction algorithm. 
We pre-train a single model on 4,140 production traces and embed it in S4-FIFO to learn optimal cache parameters. 
On 1,035 evaluation traces, S4-FIFO improves the mean efficiency by 26\% compared to S3-FIFO and by 8\% compared to 3L-Cache, the best state-of-the-art algorithm. 
S4-FIFO is also robust---increasing miss ratio over FIFO by 0.8\% on the worst trace, whereas 3L-Cache increases FIFO's miss ratio by 8.8\%. 
Finally, S4-FIFO’s decisions are also interpretable: a language model can provide a rationale for why a particular configuration was chosen.
\vspace{-1em}
\end{abstract}

\vspace{0.25em}
\section{Introduction}
\label{sec:intro}

\begin{figure}
    \centering
    \includegraphics[width=\linewidth]{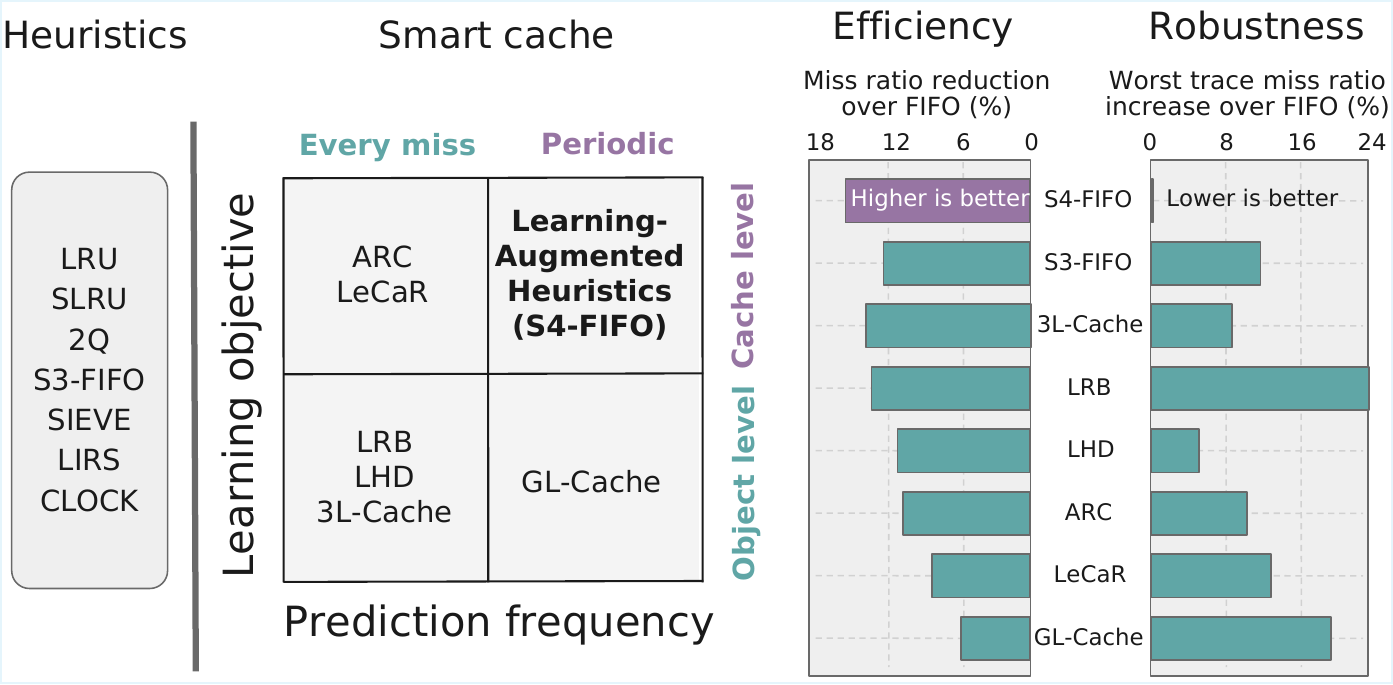}
    \vspace{-1.2em}
    \caption{A classification of eviction algorithms (\autoref{sec:motivation}) indicates that periodic cache-level learning is highly effective, yet it has not been explored. We bridge this gap with learning-augmented heuristics, exemplified by a new algorithm, \sysname, which achieves both high efficiency and strong robustness. }
    \label{fig:headline}
    \vspace{-1.8em}
\end{figure}

\vspace{-0.25em}

Caching underpins performance across modern systems---from storage stacks~\cite{cachesack,berg_cachelib_2020,shards,wu_storage_2021} and databases~\cite{2Q, DBLP:conf/usenix/SongKKPN21} to large web services~\cite{DBLP:conf/sigcomm/AtreSWB20, DBLP:conf/sigmetrics/BasuSGSS17, DBLP:conf/nsdi/BergerSH17, DBLP:conf/sigcomm/ByersCMR02, DBLP:conf/cloud/FanLAK11, DBLP:journals/tnsm/FanLLHWW21, DBLP:conf/eurosys/MokhtarianJ14, DBLP:journals/sigops/NygrenSS10, DBLP:conf/sigcomm/SchompBKMS20, DBLP:conf/conext/SundarrajanFKS17, DBLP:conf/usenix/Yan022, DBLP:conf/nsdi/YangSBRS22}---by absorbing repeated access and reducing backend load. At the heart of every cache is an eviction algorithm that decides which objects to retain under tight memory budgets. Today’s production systems overwhelmingly rely on static heuristics such as LRU, 2Q~\cite{2Q}, S3-FIFO~\cite{s3fifo}, and SIEVE~\cite{sieve} because they are simple, fast, and easy to deploy. In parallel, a growing body of work has proposed ``smart'' caches---adaptive and learned policies such as ARC~\cite{ARC}, LeCaR~\cite{LeCaR}, LRB~\cite{LRB}, LHD~\cite{LHD}, GL-Cache~\cite{GLcache}, and 3L-Cache~\cite{3L-Cache}---that aim to tailor eviction decisions to each workload. 

In principle, smart caches should dominate hand-crafted heuristics. In practice, very few of them (if any) have been adopted. Many of the smart caches learn at the granularity of individual objects and make predictions on \textit{every miss}, optimizing surrogate objectives such as reuse-distance regression loss or per-object utility scores rather than the cache’s true objective, the miss ratio. 
These object-level, per-miss learning schemes suffer from four recurring problems: (1) \textbf{objective mismatch}, where improvements in prediction metrics do not translate into fewer misses; (2) \textbf{instability and noise sensitivity}, as these schemes react to highly variable request-level signals; (3) \textbf{overheads}, due to per-object metadata, frequent inference on the critical path, and additional runtime complexity; and (4) \textbf{limited robustness and interpretability}, as eviction decisions are opaque, difficult to interpret, and may yield miss ratios higher than FIFO on some workloads.

At the same time, recent work shows that static heuristics can achieve surprisingly strong performance. S3-FIFO~\cite{s3fifo}, for example, uses a small FIFO queue, a main FIFO queue, and a metadata-only ghost queue with simple promotion rules to filter one-hit wonders and retain popular objects. 
While it is strong on the mean and median trace, our analysis reveals that its static configuration leaves substantial headroom for the tail traces. 

We classify smart eviction algorithms along two axes---\textit{learning granularity} (object vs. cache level) and \textit{prediction frequency} (per-miss vs. periodic)---and observe that existing approaches occupy three of the four quadrants (\autoref{fig:headline}). The remaining quadrant, periodic cache-level learning, remains unexplored, despite offering an appealing opportunity: directly learning cache parameters that optimize miss ratio. 

This observation motivates a fundamentally different approach to incorporating learning into caching: instead of learning a new, complex eviction policy, we propose learning to configure a simple, expressive heuristic. 
We introduce \textbf{\underline{L}}earning-\textbf{\underline{A}}ugmented \textbf{\underline{H}}euristics (LAH), a framework that cleanly separates the cache’s data and control planes. The data plane implements a deterministic heuristic with a small number of explicit knobs. The control plane runs asynchronously and infrequently, using cache-level features to select a configuration that directly targets miss-ratio reduction. A single model, pre-trained offline on a large corpus of traces, acts as a reusable ``foundation model'' for caches: it performs zero-shot prediction without per-deployment retraining. 

We realize LAH in S4-FIFO, a ``Smart S3-FIFO'' eviction algorithm that augments S3-FIFO with an additional FIFO region and exposes its internal control knobs for learning. S4-FIFO discretizes the configuration space into a small set of representative parameter combinations and trains a gradient-boosted decision tree~\cite{friedman2001greedy,xgb} to choose among them using a cost-sensitive objective that explicitly anchors robustness to FIFO. 
Trained once on 4,140 production traces spanning block, key–value, and CDN caches, this model is embedded into the algorithm as lightweight, dependency-free code. When evaluated on 1,035 held-out production traces, S4-FIFO improves mean miss-ratio reduction by 26\% over S3-FIFO and by 8\% over 3L-Cache, the best prior algorithm, while never exceeding FIFO’s miss ratio by more than 0.8\% on the worst trace. For comparison, the next-best algorithm, 3L-Cache, increases FIFO's miss ratio by 8.8\% in the worst case. Because of the separation of the control and data planes, S4-FIFO achieves throughput on par with other heuristics, such as LRU and 2Q. 
Because both its inputs and outputs are cache-level quantities with clear semantics, S4-FIFO’s decisions are also relatively interpretable: a language model can reason about the factors behind a particular configuration choice.

This paper makes the following contributions:
\begin{itemize}
    \item We introduce a cache classification and demonstrate that \textit{periodic} \textit{cache-level} learning is the most effective approach. 
    \item We propose learning-augmented heuristics (LAH), a framework that augments static heuristics with parameters and uses a foundation model to learn them.
    \item We design and implement \sysname, an example of LAH. We pre-train a gradient boosting tree model for \sysname on 4,140 production traces and provide a dependency-free library for users~\footnote{\url{https://libcachesim.com}}. 
    \item We evaluate \sysname on 1,035 production traces and show that it is both more efficient and more robust than state-of-the-art algorithms, while matching the throughput of heuristics. Moreover, S4-FIFO’s decisions are interpretable. 
\end{itemize}



\vspace{-1em}
\section{Background and Motivations}
\vspace{-0.5em}
\label{sec:bg}

\vspace{-0.25em}
\subsection{Evolution of Cache Eviction Algorithms}
\vspace{-0.5em}
A central component of cache performance is the eviction algorithm, deciding which object(s) to remove when space runs out. Over the years, researchers and practitioners have designed a wide spectrum of eviction strategies, from simple heuristics to learning-based policies. 

\vspace{-0.75em}
\subsubsection{Heuristics-Based Algorithms}
\vspace{-0.5em}
Most (if not all) production systems use heuristics-based eviction algorithms today. 
Traditional heuristics, such as 2Q~\cite{2Q} and LIRS~\cite{LIRS}, and modern heuristics, such as S3-FIFO~\cite{s3fifo} and SIEVE~\cite{sieve}, operate based on \textit{manually} curated rules that exploit recency and frequency patterns in a workload. 
These heuristics are valued for their simplicity, high throughput, and ease of implementation. 
However, they often fall short in efficiency compared to adaptive algorithms. 

\vspace{-0.5em}
\subsubsection{Smart Cache Eviction Algorithms}
\vspace{-0.5em}
\paragraph{Adaptive algorithms.}
For example, ARC~\cite{ARC} maintains four LRU queues: two for data and two for ghost. It dynamically resizes recency and frequency data queues based on observed hits on the corresponding ghost queues.
Similarly, DLIRS~\cite{li2018dlirs} extends LIRS by introducing an adaptive partition between low-reuse (HIR) and high-reuse (LIR) regions, which is dynamically resized based on its reuse distance. Although adaptive cache eviction algorithms do not employ machine learning models, we classify them as smart cache eviction algorithms because they can adapt to workloads by observing dynamic features. 

\vspace{-0.5em}
\paragraph{Learning-based algorithms.}
More recent work builds prediction models to guide eviction decisions. Algorithms such as LRB~\cite{LRB}, 3L-Cache~\cite{3L-Cache}, and GL-Cache~\cite{GLcache}  use supervised learning at the object level to estimate reuse distance and then evict the object predicted to have the least future value. Other approaches like LHD~\cite{LHD} model eviction using conditional hit-density, while LeCaR ~\cite{LeCaR} adaptively reweights LFU and LRU weights with regret minimization. Although these complex policies may outperform heuristics, deploying them adds practical challenges, mainly: (1) online model training and inference add overheads, reducing cache throughput, and (2) their decision-making logic is opaque, from their features to their outputs. As a result, it is hard to interpret, tune, or debug why they behaved the way they did.

\begin{table*}[]
\footnotesize
\centering
\caption{Classifying smart caches based on learning granularity and prediction frequency. \vspace{-0.8em}}
\label{table:classification}
\begin{tabular}{@{}llll@{}}
\toprule
{Learning granularity} & {Prediction frequency} & {Example }            & {Problems}          \\ \midrule
\multirow{2}{*}{Object-level} & Every miss & LRB~\cite{LRB}, LHD~\cite{LHD}, 3L-Cache~\cite{3L-Cache}  & Objective mismatch, unpredictability, overheads \\ \cmidrule(l){2-4} 
                     & Periodic         & GL-Cache~\cite{GLcache}           & Stale prediction  \\ \midrule
\multirow{2}{*}{Cache-level}  & Every miss & ARC~\cite{ARC}, LeCaR~\cite{LeCaR} & Instability, delayed reward                     \\ \cmidrule(l){2-4} 
                     & Periodic         & S4-FIFO (this work) & None of the above \\ \bottomrule
\end{tabular}
\vspace{-0.8em}
\end{table*}

\begin{figure}[t]
    \centering
    \includegraphics[width=0.96\linewidth]{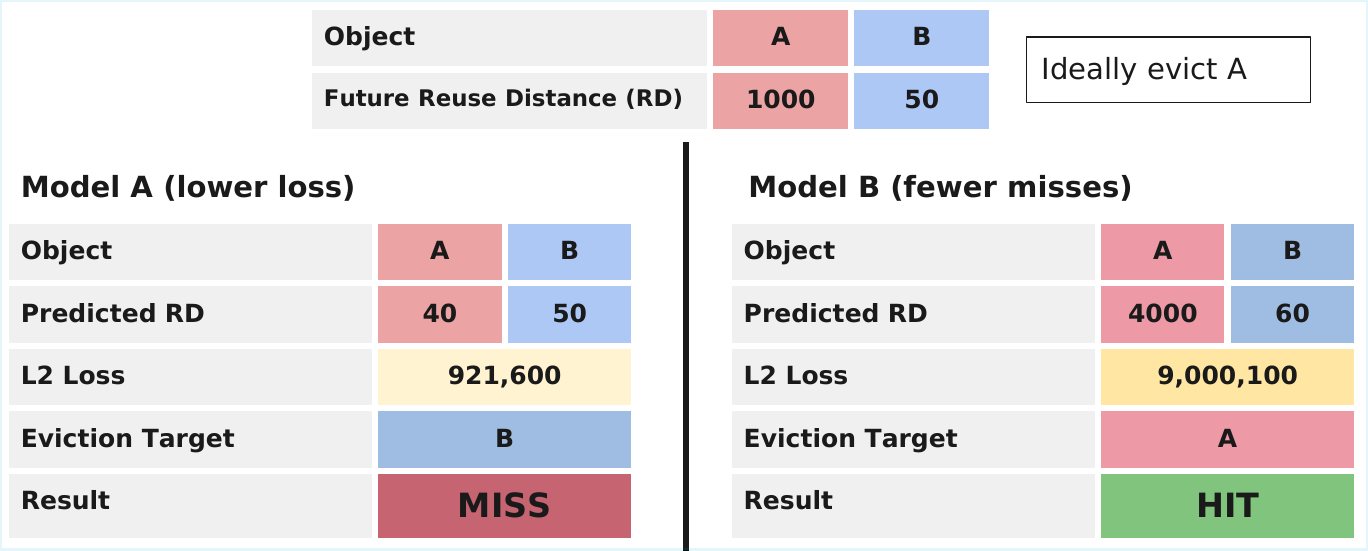}
    \caption{Object-level learning has an objective mismatch issue. A smaller loss on object-level metrics, e.g., reuse distance, does not guarantee fewer cache misses.}
    \label{fig:objective_mismatch}
    \vspace{-1.6em}
\end{figure}

\subsection{Classifying Smart Eviction Algorithms}
\label{sec:motivation}
We classify existing smart cache eviction algorithms along two dimensions: the \emph{granularity of learning} and the \emph{frequency of prediction or adaptation} (\autoref{table:classification}).

\paragraph{Learning granularity.} A smart cache eviction algorithm adapts to workloads by making predictions about the workload. 
These predictions can be either at the object level or the cache level. \emph{Object-level algorithms}, such as 3L-Cache~\cite{3L-Cache}, LRB~\cite{LRB}, and LHD ~\cite{LHD} assign utility score or eviction probability to every individual object in the cache. For example, LRB and 3L-Cache learn the reuse distance for objects in the cache to predict the next access. Note that we also classify GL-Cache~\cite{GLcache} as an object-level algorithm because it makes predictions for object groups similar to objects. 

In contrast, \emph{cache-level algorithms} learn by optimizing the internal cache parameters rather than evaluating individual objects. A classic example is ARC~\cite{ARC}, which dynamically tunes the target sizes of its recency and frequency queues based on workload patterns, without assigning scores to cached objects. Similarly, LeCaR~\cite{LeCaR} dynamically adjusts the expert weights based on hits on the ghost queues. 

\paragraph{Prediction frequency.} Besides granularity, the frequency at which an algorithm predicts or adapts is also critical. Most prior algorithms (e.g., LRB, ARC, LeCaR) make a prediction \emph{every miss}. For example, LRB samples 64 objects and predicts their reuse distance at each eviction; ARC updates the queue sizes upon each hit on the ghost queue (cache miss).

GL-Cache is the only algorithm that makes predictions \emph{periodically} by ranking all object groups and evicting the top 10\%. It does not trigger another prediction before all the top 10\% object groups are evicted. 


\subsection{Learning Granularity: Object vs Cache}
\subsubsection{Object-level Learning}

Several recent approaches operate at the granularity of individual objects, typically by predicting reuse distance or popularity. While this formulation appears natural, object-level learning faces inherent challenges regarding \emph{objective alignment}, \emph{predictability}, and \emph{robustness and interpretability}. 

\paragraph{Objective mismatch.} The first challenge with object-level learning is objective mismatch, a phenomenon akin to Goodhart's Law: when a proxy becomes the target, it ceases to be a good measure. Many learned caches optimize for object-level metrics (e.g., minimizing L2 loss on reuse distance) rather than the actual miss ratio objective. For example, \autoref{fig:objective_mismatch} shows an example where model A achieves a lower L2 loss than model B, yet it fails to evict the correct object. 




\paragraph{Unpredictability.} Object-level metrics often cannot be predicted accurately. First, metrics, such as object reuse distance, are not intrinsic to an object and are heavily influenced by the workload, such as burstiness. 
Furthermore, cache workloads often exhibit a large number of one-hit wonders~\cite{s3fifo}, for which no information (e.g., past $N$ reuse distance, frequency) is available for learning. Some object-level metrics, such as popularity, are intrinsic properties of an object; however, they change quickly over time and cannot be used directly for eviction. 

\paragraph{Robustness and interpretability.}
As object-level learning is driven by the unpredictable metrics described above, it becomes highly sensitive. Variations in workload can produce wrong eviction decisions. This volatility directly translates into poor robustness; models may perform well on average yet fail badly on certain workloads. 
Moreover, the outputs of object-level learning are fundamentally opaque. The model assigns per-object scores with little to no operational semantics, leaving operators unable to explain when the miss ratio increases. This combination of unpredictable inputs and opaque outputs makes object-level learning risky to deploy at scale.


\subsubsection{Cache-level Learning}
Cache-level learning avoids the aforementioned problems by tuning a small set of global cache parameters. 
For example, ARC~\cite{ARC} maintains two data queues and two ghost queues. Data queues store data, while ghost queues only keep metadata of recently evicted objects. ARC dynamically adjusts the queue size based on ghost hits. Intuitively, the more hits on the ghost queue, the more space is needed for the corresponding data queue. 
As another example, LeCaR~\cite{LeCaR} uses two eviction algorithms as experts and randomly chooses one expert to pick an eviction candidate based on its weight. The expert (algorithm) with a larger weight is more likely to be used for eviction. Similar to ARC, LeCaR adjusts experts' weights based on ghost hits using reinforcement learning. 

\vspace{-0.5em}
\paragraph{Direct optimization.} \textit{Cache-level learning often directly optimizes for the miss ratio}. For example, a ghost hit means that if the data queue is larger, it will be a cache hit, so increasing the data queue size directly reduces misses. Similarly, if an expert has many ghost hits in LeCaR, indicating it is not suitable for the workloads, then reducing its weight would reduce the number of wrong evictions. Direct optimization not only helps reduce the miss ratio but also improves interpretability, enabling operators to reason about specific actions. 

\vspace{-0.5em}
\paragraph{Magic parameters.} Although cache-level learning is intuitive and avoids the objective-mismatch problem, it often has magic parameters. For example, how much and how often the parameters should be updated. Previous work has shown that the default move \emph{one} slot \emph{per ghost hit} approach in ARC has pathological failure cases~\cite{s3fifo}. 

\begin{figure}[t]
  \centering
    \begin{subfigure}[t]{0.48\linewidth}
    \centering
    \includegraphics[width=\linewidth]{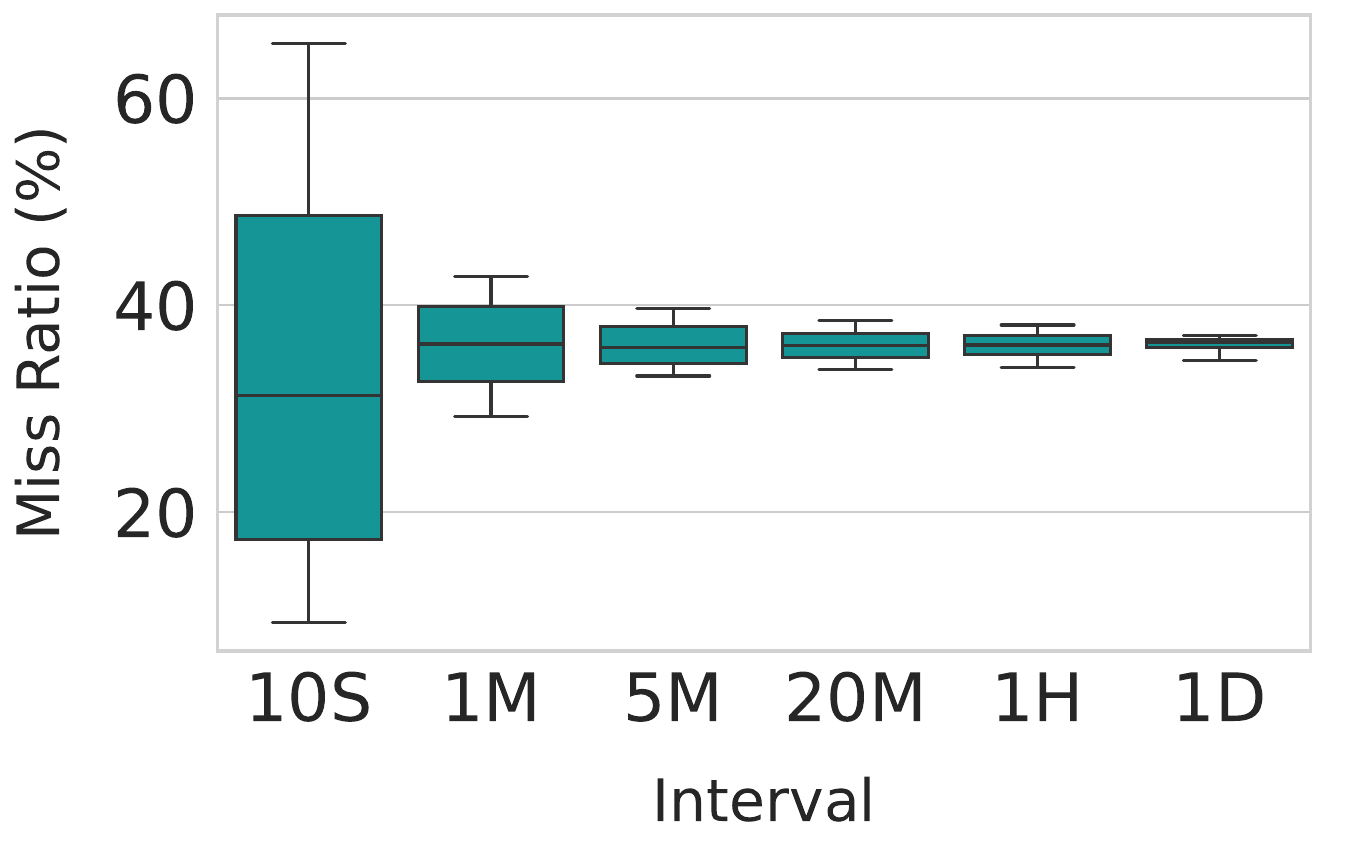}
    \end{subfigure}
    \hfill
    \begin{subfigure}[t]{0.48\linewidth}
    \centering
    \includegraphics[width=\linewidth]{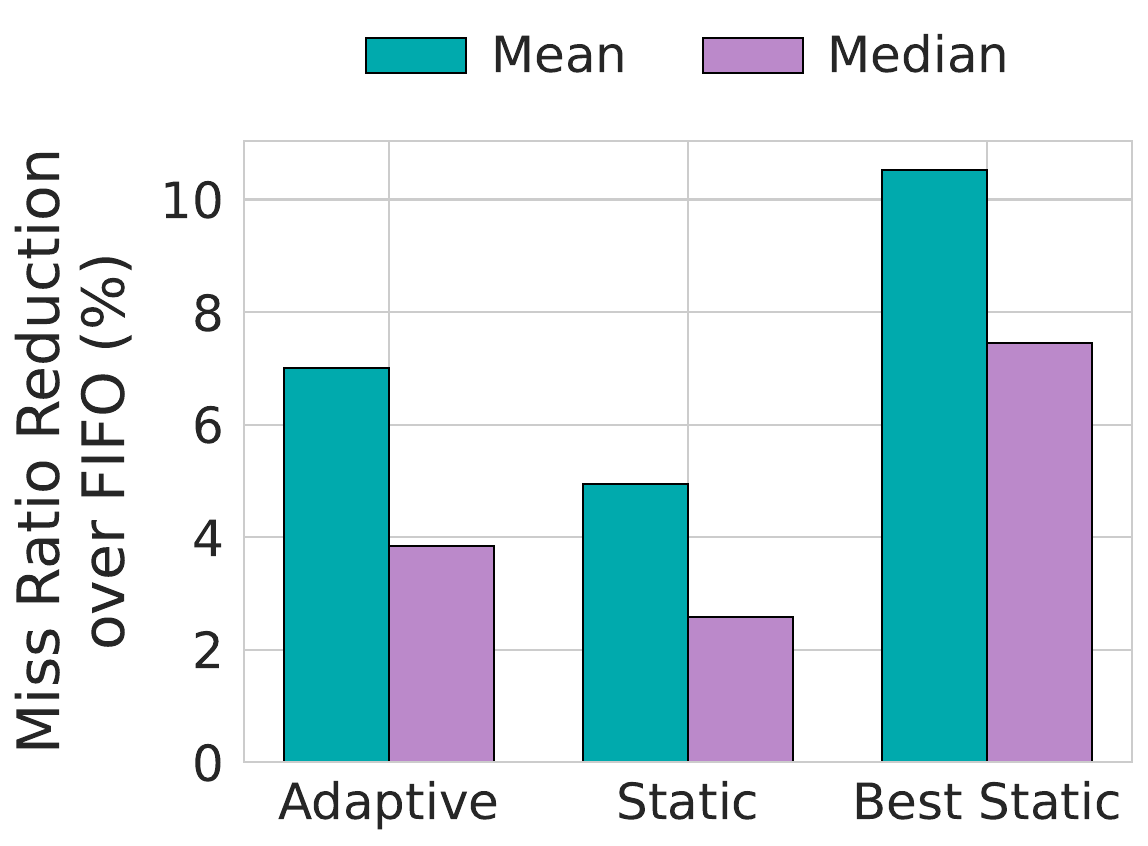}
  \end{subfigure}
  \vspace{-0.5em}
  \caption{(a) Cache metrics, such as miss ratio, exhibit greater variance at finer time granularity. (b) Choosing the correct parameters can outperform LeCaR's per-miss adaptivity.
  \vspace{-1.6em}}
  \label{fig:motivation:permiss}  
\end{figure}


\subsection{Prediction Frequency: Per-miss vs Periodic}

While the learning objective is critical for the effectiveness of a smart cache, how often a cache makes a prediction or adaptation is equally important. 

\subsubsection{Per-miss Prediction and Adaptation}
Most algorithms make predictions at the request granularity, e.g., every miss. 
For example, LRB estimates object reuse distance on each cache miss, while ARC and LeCaR update queue sizes or weights on every hit to their ghost queues. Although such per-miss adaptation can appear highly responsive, it suffers from three key limitations: \emph{instability and noise overfitting}, \emph{delayed reward}, and \emph{inference overhead}. 

\paragraph{Instability and overfitting to noise. }
Observations on each request and object might not reflect the overall patterns in the workload. By reacting to every miss, the algorithms will instead ``chase'' this noise. We find that real-world workloads often exhibit significant variance at fine time granularity, and request-level signals are often very noisy. \autoref{fig:motivation:permiss} left shows the miss ratio measured at different time granularities on a CloudPhysics trace. We observe that at smaller time intervals, the miss ratio over time tends to have a very large variance. 

We further measured the impact of per-miss adaptivity using 106 traces from the CloudPhysics dataset. \autoref{fig:motivation:permiss} right shows that, although per-miss adaptivity improves over using a single static parameter for \textit{all} traces (static), it is significantly worse than choosing the best static parameter for each trace. This suggests that choosing the right parameter is much more important than per-miss adaptivity. 

\vspace{-0.5em}
\paragraph{Delayed Reward for Cache-level Learning. }
Besides overfitting, per-miss learning also suffers from the delayed reward problem. Because a cache miss may happen \emph{long} after an object is evicted, and parameter changes take time to become effective, we find that the change often lags behind the need. This lag makes the feedback signal slow and ambiguous. 


\vspace{-0.5em}
\paragraph{Overheads for Object-level Learning. }
When pairing per-miss prediction with object-level learning, another drawback is the overhead. Performing a model inference on each miss significantly reduces the throughput. Besides, object-level learning also incurs significant per-object metadata storage overhead. For example, LRB adds over 200 bytes of features to each object, which eats into DRAM capacity. 

\subsubsection{Periodic Prediction}
To the best of our knowledge, GL-Cache~\cite{GLcache} is the only cache eviction algorithm that uses periodic prediction. Similar to other object-level learned caches, GL-Cache predicts a score for each object group. However, instead of predicting on the critical path, GL-Cache makes one prediction \emph{periodically} and reuses the scores until 10\% of the object groups are evicted. However, because the usefulness of objects and groups changes over time, their predictions quickly become stale, and periodic prediction often leads to lower efficiency when used with object-level learning. 

Our analysis of existing algorithms reveals a dichotomy in the design space. Object-level algorithms suffer from objective mismatch, unpredictability, and huge overheads. Moreover, per-miss learning algorithms are prone to instability and delayed rewards.
\autoref{table:classification} highlights a distinct gap in the design space---we need smart cache eviction algorithms that \emph{learn and adapt at the cache level, with the learning objective aligned with miss ratio reduction}; meanwhile, we need \emph{periodic learning} to avoid unnecessary overfitting to noise and magic parameters. 



\begin{figure*}[t]
    \centering
    \includegraphics[width=0.96\linewidth]{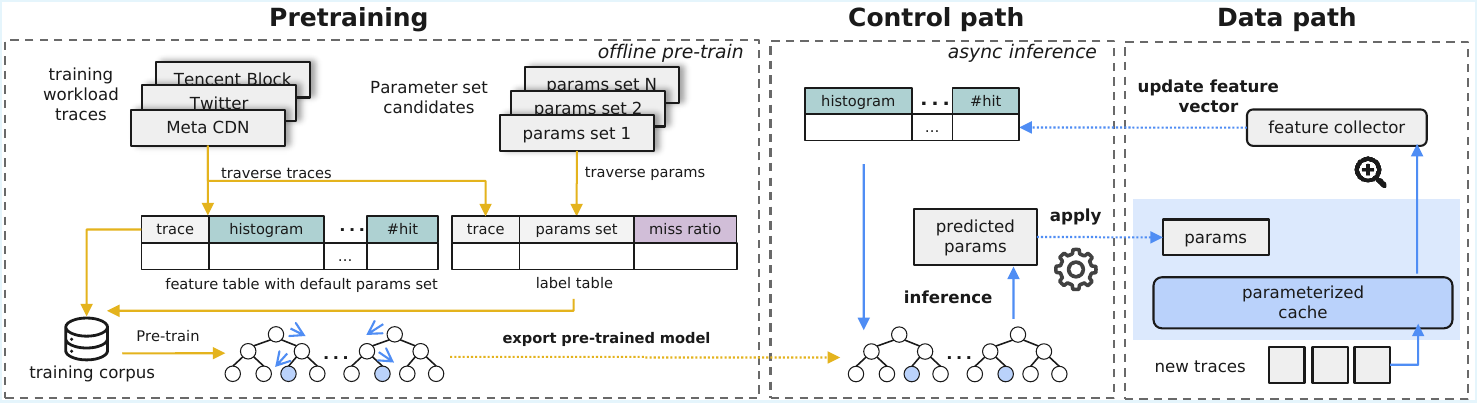}
    \caption{Illustration of learning-augmented heuristics. (1) Offline pretraining (left): A foundation model is trained on a massive corpus of trace features and optimal parameter labels to learn universal caching rules. 
    (2) Online operation (right): The system decouples the critical \textit{data path} from the \textit{control path}. The data path collects lightweight workload-level features, while the control path asynchronously queries the pre-trained model to adapt the cache parameters to the current workload trace.}
    \vspace{-1.25em}
    \label{fig:lah}
\end{figure*}

\section{Learning-Augmented Heuristics (LAH)}
\label{sec:lah}
In this section, we introduce a \textit{learning-to-configure} paradigm that fundamentally rethinks how machine learning interacts with caching systems. Instead of embedding complex learning logic (e.g., neural networks) directly into the critical path, we propose a clean separation of concerns: \emph{the cache data path executes a simple, parameterized heuristic, while the control plane utilizes a pre-trained ``foundation model'' to choose the heuristic's parameters}, an approach we call Learning-Augmented Heuristics (\emph{LAH}). 

Existing learning-based caches (e.g., GL-Cache, LeCaR) often treat learning as an online, instance-specific task and train from scratch for each new or even the same workload. This approach suffers from ``catastrophic forgetting'' and high retraining costs. 
In contrast, our approach builds upon the well-established insight that cache access patterns are governed by universal characteristics: phenomena like scanning, looping, and thrashing are structural behaviors that transcend specific datasets~\cite{workingset, LIRS}. By extracting \textit{content-oblivious} features rather than tracking specific object IDs, we can train a single global model that generalizes across diverse workloads. 
This effectively builds a {foundation model for caching}: knowledge learned from thousands of offline traces is preserved and transferred to new, unseen environments. The model does not need to relearn but recognizes the pattern and applies the optimal configurations. 
This decoupled architecture offers five critical advantages over tightly coupled learned caches. 

\vspace{-0.5em}
\paragraph{Simplicity.} The data path logic remains simple and deterministic. It requires almost no metadata overhead (unlike LRB or 3L-Cache, which track extensive ghost histories per object) and avoids complex data structures. This ``stateless'' nature drastically reduces code complexity and bug surface area, making the system easier to verify and deploy in production.

\paragraph{Performance.} By offloading the inference to an asynchronous control plane, the critical path for {GET} and {PUT} operations remains extremely lightweight, 
ensuring high throughput and low tail latency. 

\paragraph{Efficiency.} Simple static heuristics, such as 2Q, cannot achieve high efficiency compared to state-of-the-art algorithms. LAH can automatically optimize for different workloads. During the cache warm-up period, LAH identifies the ``golden configurations'' and updates its parameters to match the workload's patterns. 

\vspace{-0.75em}
\paragraph{Robustness.} When evaluating state-of-the-art eviction algorithms on thousands of production workloads, we find that most of them exhibit poor robustness---some traces exhibit a miss ratio significantly higher than a simple baseline such as FIFO. This hinders adoption because users do not know whether these algorithms will work on their workloads. 
LAH learns the configuration from high-level workload features, which are less susceptible to perturbations and noise. Moreover, LAH selects parameters from a validated set of safe configurations. These enable LAH to achieve robustness. 


\vspace{-0.75em}
\paragraph{Interpretability.}
LAH operates on aggregated metrics such as hit distributions, which align naturally with how operators already think about and understand cache behavior. 
These high-level signals are then mapped to a small set of heuristic parameters, so each prediction corresponds to a knob with a clear operational meaning and existing tuning intuition. 
Because both the inputs and outputs live in this shared, cache-centric space, the resulting configurations are much more interpretable than existing learned caches. 

While LAH can be designed on top of different parameterized algorithms, we use S3-FIFO~\cite{s3fifo} as an example because it is simple, performant, and scalable. 
In the next section, we describe \sysname, a new cache eviction algorithm that is \underline{S}imple, \underline{S}calable, \underline{S}mart, and only uses \underline{S}tatic FIFO queues. 
\sysname exposes and augments S3-FIFO's internal parameters to our foundation model for learning, transforming a static heuristic into an efficient, robust, and interpretable learning-augmented heuristic.

\section{\sysname Design and Implementation}
\vspace{-0.5em}
\label{sec:design}
In this section, we describe the design of \sysname, detailing the heuristic data path, feature engineering, configuration quantization, and model training. 

\vspace{-1em}
\subsection{Overview}
We illustrate the high-level overview as shown in \autoref{fig:lah}. 
\vspace{-0.5em}

\paragraph{Heuristic Augmentation.} 
We first augment S3-FIFO with a burst-aware parameter that virtually partitions the small queue into an additional FIFO queue, giving the parameterized heuristic greater flexibility to minimize the miss ratio.

\paragraph{Dataset Preparation. } 
We randomly split the 5,175 traces used in this study into training and test sets to avoid data leakage. 
Then we perform a grid search to determine the best parameters (labels for training) for each trace. 
Next, we collect cache and workload-level features by running \sysname using the default parameter on each trace, yielding the samples used to train the model. 

\paragraph{Model Pretraining. }
We use features and labels from the previous step to train a classification model that predicts the best parameters for an unseen workload. 

\subsection{Augmenting the Parameters}
\begin{figure}[t]
    \centering
    \includegraphics[width=0.96\linewidth]{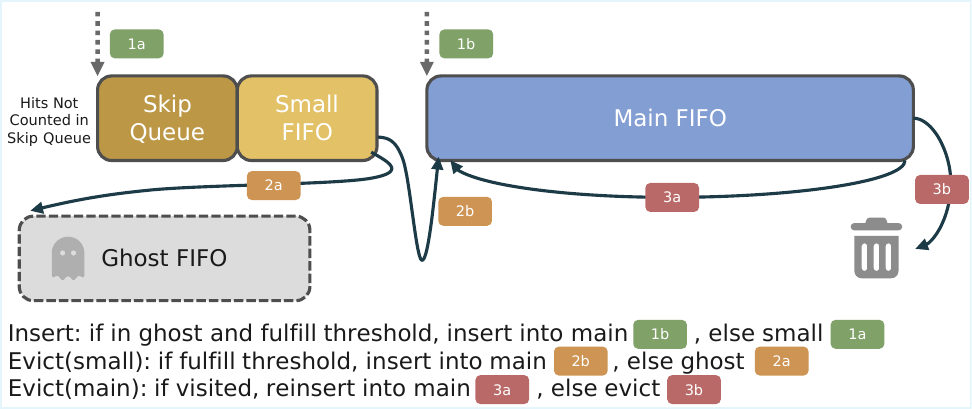}
    \vspace{-0.5em}
    \caption{An illustration of the data path in \sysname.}
    \vspace{-0.5em}
    \label{fig:data-path}
\end{figure}

\paragraph{Parameter prediction and request serving.}
The previous steps only need to be performed once offline. Once the model is trained, it will be used in different settings without retraining. 
When a cache starts, \sysname collects features online and then asynchronously makes a single prediction. Then \sysname switches from the default parameters to the predicted parameters. \rev{Queue resizing is lazy: when the predicted configuration reduces a queue's target size, \sysname does not immediately move or evict objects. Instead, future evictions preferentially come from queues that exceed their new limits until the queue sizes converge. This allows parameter updates without pausing the cache or adding extra work to the request critical path.} \revv{
For simplicity, our evaluation makes a single prediction per trace. We found that substantial shifts requiring a new prediction are uncommon. In practice, \sysname could be made to periodically re-predict its parameters if needed, for example on a weekly basis. We leave the choice of refresh policy and interval outside the scope of this work.
}


S3-FIFO maintains three FIFO queues. A small queue, occupying 10\% of the cache, filters out one-hit wonders; a main queue retains popular objects via reinsertion; and a metadata-only ghost queue identifies moderately popular objects that were mistakenly evicted from the small queue.
Each cached object is associated with a 2-bit saturated frequency counter that is incremented on every cache hit, and a threshold on this counter governs movement between the queues. 
\rev{\sysname preserves the same three physical FIFO queues, but parameterizes their behavior, as shown in \autoref{fig:data-path}.}
The parameters are shown in \autoref{table:parameter}.

\vspace{-0.5em}
\paragraph{Queue size.} 
While S3-FIFO sets the small queue at 10\% of the cache, \sysname treats the queue sizes as tunable parameters. This allows the system to select the optimal ratio between the small and protected regions based on workload characteristics. In addition, \sysname also makes the ghost queue size tunable to collect more information about the workload. 

\vspace{-0.5em}
\paragraph{Promotion threshold. } 
S3-FIFO moves an object evicted from the small queue to the main queue if the object's frequency is greater than 1. \sysname makes the frequency threshold learnable, allowing control over how strictly the small queue filters new insertions. Similarly, to handle cyclic scan patterns~\cite{LIRS, LRU-K}, \sysname learns the frequency threshold for promoting an object from the ghost queue to the main queue.

\vspace{-0.5em}
\paragraph{Skip ratio. }
To mitigate correlated references (bursty ``one-hit wonders'')~\cite{2Q, Clock2Q+}, \sysname does not increment the frequency counter for objects in the first $\kappa$ fraction of the small queue. 
\rev{This creates a virtual probationary region inside the small queue, adding an extra filtering stage without introducing a fourth physical FIFO queue.}


\begin{table}[t]
    \scriptsize 
    \centering
    \caption{Hyperparameters of the \sysname.}
    \vspace{-1em}
    \label{table:parameter}
    \begin{tabular}{clcc} 
    \toprule
        \textbf{Symbol} & \textbf{Physical Meaning} & \textbf{Default} & \textbf{Learning range} \\
    \midrule
        $\rho_{\mathcal{S}}$      & Small queue size ratio    & $0.10$ & $[0.05, 0.90]$ \\
        $\rho_{\mathcal{G}}$      & Ghost queue size ratio    & $0.90$ & $[0.90, 6.00]$ \\
        $\kappa $ & Small skip ratio & $0.00$ & $\{0.00, 0.25\}$ \\
        $\tau_{\mathcal{S}}$      & Small-to-Main promotion threshold & $2$    & $\{1, 2\}$ \\
        $\tau_{\mathcal{G}}$      & Ghost-to-Main promotion threshold & $0$    & $\{0, 1\}$ \\
    \bottomrule
    \end{tabular}
    \vspace{-1.75em}
\end{table}

\paragraph{Parameter discretization. }
The parameter space of \sysname is continuous and high-dimensional, and its exponentially growing combinations make it impossible to perform a grid search to find the best parameters. 
Therefore, we discretize the parameter ranges and choose a few values for each based on domain knowledge. 
Specifically, the small queue size can take up 5\%, 10\%, 20\%, 30\%, 50\%, 70\%, or 90\% of cache capacity; the ghost queue can store the same number of objects, $3\times$ or $6\times$ that of the main queue; the small-to-main promotion threshold can be either 1 or 2; the ghost-to-main threshold can be either 0 or 1; and the skip ratio can be either 0 or 0.25.


\subsection{Learning the Parameters}
\subsubsection{Learning Objective}\label{subsec:learningObj}

To enable efficient learning, we formulate the parameter learning as a probabilistic classification task over a finite set of candidate parameter configurations $\mathcal{C}$. We learn a conditional probability distribution $P(y=c_j | \mathbf{x}; \phi)$ over these candidates $c_j \in \mathcal{C}$ given the workload features $\mathbf{x}$. 


Standard classification objectives (e.g., cross-entropy) treat all misclassifications symmetrically. However, in caching, error costs are highly \textit{asymmetric and pairwise}: mistaking a scan-heavy workload for a recency-friendly one causes catastrophic thrashing, whereas the reverse error often yields only marginal degradation. 
To capture these varying penalties, we employ a data-driven cost matrix $\mathbf{L} \in \mathbb{R}^{K \times K}$, where $K$ is the number of candidate parameter configurations. Each entry $L_{kj}$ quantifies the pairwise ``regret'' of selecting configuration $c_k$ when the true optimal is $c_j$, normalized by the miss ratio of a baseline anchor $\mathcal{M}_{\text{anchor}}$,

\begin{equation*} 
    L_{kj} = \mathbb{E}_{\text{train}} \left[ \frac{\text{MR}(c_k) - \text{MR}(c_j)}{\mathcal{M}_{\text{anchor}}} \right].
\end{equation*}

\paragraph{Choice of Anchor.} While the anchor can theoretically be any baseline (e.g., LRU, FIFO), we select $\mathcal{M}_{\text{anchor}} = \text{MR}_{\text{FIFO}}$. This choice is motivated by the structural nature of our system: (1) \sysname is fundamentally built upon FIFO queues, therefore FIFO serves as the natural lower bound for robustness, and (2) normalizing by FIFO makes the cost metric invariant to the cacheability of the trace, allowing the model to learn effectively across workloads with vastly different miss ratios.

During inference, we minimize the expected risk,
\begin{equation*} 
    \hat{c} = \arg\min_{c_k \in \mathcal{C}} \sum_{j} P_\phi(c_j | \mathbf{x}) \cdot L_{kj}.
\end{equation*}

\vspace{-2.5 em}
\subsubsection{Features}
\begin{table}
    \scriptsize
    \centering
    \caption{Features used by \sysname.}
    \vspace{-1em}
    \label{table:feature}
    \begin{tabular}{ccc}
    \toprule
        Features & Dimension & Meaning \\
        \midrule 
        Small histogram & 20  &  Hit-position distribution ratio in $\mathcal{S}$ \\
        Main histogram & 20  &  Hit-position distribution ratio in $\mathcal{M}$ \\
        Ghost histogram & 20 &  Hit-position distribution ratio in $\mathcal{G}$ \\

        Small hits ratio & 1 & Proportion of hits happening in $\mathcal{S}$  \\
        Main hits ratio & 1 & Proportion of hits happening in $\mathcal{M}$ \\
        Ghost hits ratio & 1 &  Proportion of hits happening in $\mathcal{G}$ \\
        Log cache size & 1 & log(cache size)\\
        Utility gap & 1 & Utility divergence between $\mathcal{S}$ and $\mathcal{M}$ \\
        Filtering efficiency & 1 & Relative hit density of $\mathcal{S}$ vs. $\mathcal{M}$ \\
        Ghost pressure & 1 & Intensity of miss-on-eviction events \\
        Tail heaviness & 1 & Hit concentration at the deep end of $\mathcal{M}$ \\
        Decay rate & 1 & Recency decay slope at the head of $\mathcal{S}$ \\
        One-hit ratio & 1 & Estimated fraction of one-hit wonders \\
        Unique ratio & 1 & Workload churn rate (unique objects ratio) \\
        Scan intensity & 1 & Effective scan pressure under capacity limits \\
        Thrashing risk & 1 & Signal for working set exceeding capacity \\
    \bottomrule
    \end{tabular}
    \vspace{-2em}
\end{table}

To enable lightweight yet effective learning, we model the cache states and workload characteristics $s_{t}$ using a compact feature vector (73 dimensions). The list of features used can be found in \autoref{table:feature}, where $\mathcal{S}$, $\mathcal{M}$, and $\mathcal{G}$ are the small queue, the main queue, and the ghost queue, respectively.

\paragraph{Cache metrics and workload characteristics.} 
Cache size is critical for scale-invariant learning: the effectiveness of the small queue depends on its \emph{absolute} capacity to absorb bursts, not just its relative share of the cache. For example, a 10\% partition in a 100\,GB cache provides ample buffer space, whereas 10\% of a 100\,MB cache may be insufficient for the same workload. Therefore, we track cache size using the log-transformed capacity.
To capture the ``shape'' of locality beyond simple scalar averages, we maintain lightweight histograms (20 equally-sized bins each) for the small, main, and ghost queues. These histograms capture the distribution of hit position within each queue. For instance, the first bin stores the number of hits in the first 5\% of the queue. Intuitively, the histograms enable the cache to learn the queue usage. For example, many ghost hits suggest the small queue might be too small. 
Besides the histogram, we also include the number of hits per queue and the total number of requests. 


\vspace{-1.0em}
\paragraph{Composite features.} 
To avoid complex models, we use feature crossing to capture non-linear relationships between features, following common practice in traditional recommendation systems that rely on simple models~\cite{FM,RecFM}.
\begin{itemize}
\item \textit{Utility gap} measures the utility divergence between the small and main queues and is defined as the absolute hit difference between them divided by total hits, $\frac{H_{\text{main}} - H_{\text{small}}}{H_{total}}$. A positive gap indicates that the main queue is effectively retaining hotter items than the small queue.
\item \textit{Filtering efficiency} quantifies how well the small queue filters one-hit, computed as the ratio of hits in the small queue to those in the main queue, $\frac{H_{\text{small}}}{H_{\text{main}}}$.
\item \textit{Ghost pressure} measures the fraction of hits served by the ghost queue, $\frac{H_{\text{ghost}}}{H_{\text{ghost}} + H_{\text{total}}}$. A high ghost pressure suggests that the small queue is undersized or that the promotion threshold is too conservative.
\item \textit{Tail heaviness} measures the dependence of the workload on long-term retention and is computed as the sum of hits in the last 10 bins of the main queue’s histogram, with larger values indicating that a substantial fraction of hits comes from long-resident objects.
\item \textit{Decay rate} measures how quickly items lose utility in the probationary small queue and is computed as the slope between the first two bins of the small queue’s histogram, where a steep negative slope indicates that most objects become cold quickly. 
\item \textit{Unique object ratio} estimates cacheability and is computed as $\frac{\text{\# insertions to small}}{\text{\# requests}}$. 
\item \textit{One-hit ratio} estimates the fraction of one-hit wonders in the workload and is computed as
$\rho_{\text{onehit}} = \frac{N_{\text{onehit}}}{N_{\text{unique}}}$,
where $N_{\text{onehit}}$ is the number of objects evicted from the small queue without reaching the promotion threshold, and $N_{\text{unique}}$ is the number of unique objects observed during the window.
\item \textit{Scan intensity} and \textit{thrashing risk} are derived by normalizing \textit{unique object ratio} and \textit{one-hit ratio} by the request rate, providing the model with explicit signals to distinguish between benign scans and destructive thrashing. 
\end{itemize}

\subsubsection{Model and learning}
\paragraph{Inference Model.}
We employ a Gradient Boosting Decision Tree ensemble as the predictive model $\phi$. It accepts the 73-dimensional feature vector $\mathbf{x}$ and outputs a probability distribution over the 18 classes. We choose GBDT over deep neural networks (DNNs) for two reasons. First, it can be easily packaged into a few portable headers for distribution, allowing users to use the pre-trained model in other programming languages without installing any new libraries. For example, our Cachelib~\cite{berg_cachelib_2020} prototype employs the same model we trained using the libCacheSim~\cite{libCacheSim} simulator. 
Second, GBDT excels at handling heterogeneous, nonlinear features (e.g., histograms mixed with scalar ratios) and needs no complex normalization. 

\paragraph{Parameter search space reduction. }
Although we have discretized each parameter into a few options, the number of possible parameter combinations remains huge, which reduces learning effectiveness. 
We employ a clustering-based strategy to reduce the search space. We use the offline grid search results to identify the most common parameter combinations and identify 18 representative parameter sets out of 168 candidates via greedy set cover algorithm~\cite{GreedySetCover}.

\subsection{Implementation}
\label{sec:design:implementation}
We implement \sysname in both libCacheSim for simulation-based evaluation and Meta Cachelib for prototype evaluation and production deployment. Both implementations extend S3-FIFO and employ the same pre-trained model, trained on features and labels collected in the simulator. In each system, a hash table is used for indexing, and linked lists are used to implement the queues. 
\subsubsection{Model implementation}
We use LightGBM for our model implementation in the simulator. Because tree-based models can be compiled into sequences of conditional branches, we export the pretrained model into simple, \textit{dependency-free} libraries using m2cgen~\cite{m2cgen}, targeting several common programming languages, including C/C++, Go, Rust, Java, and JavaScript. This enables users to use our model directly without relying on LightGBM or other machine learning libraries. For example, our Cachelib implementation uses the C++ header files. 

\subsubsection{Feature collection}
Most of the features are simple counters and straightforward to maintain, so we focus on the histogram features. 

\paragraph{Small and Main FIFO queues.} We use insertion time (i.e., the insertion index) to calculate the histogram bin for each object as $\lceil \frac{T_{obj} - T_{tail}}{w} \rceil$ where $T$ denotes the insertion time and $w$ is the bin width. For example, suppose the queue has a capacity of 1000, and we divide it into 20 bins, each with a width of 50. If the tail object was inserted at time $T_{tail}=20$ and the current object at $T_{obj}=80$, then the object should be incrementing the counter at the $\lceil \frac{80-20}{50} \rceil =2^{nd}$ bin. 

\paragraph{Ghost queue.}
Computing histogram bin indices in the ghost queue is more involved because a hit in the ghost queue removes the corresponding metadata entry and shifts all subsequent entries forward. As a result, using insertion time alone to determine the bin index yields incorrect positions. To compensate, we maintain an additional \emph{deletion-count} histogram alongside the hit-position histogram for the ghost queue.
Concretely, let $i$ be the bin index obtained from the insertion-time-based formula. We sum the deletions recorded in bins 1 through $i-1$, divide this sum by the bin width $w$ to calculate the number of shifted bins $s$, and then correct the bin index as $i' = i-s$. This adjusted index is used as the hit-position bin in the ghost-queue histogram. 
Even with the adjustment, the calculated bin is still approximate because we do not consider the deleted objects in bin $i$. However, the approximation has bounded error and is at most off by 1. 

\subsubsection{Overhead Analysis}
\label{sec:design:overheads} 
We analyze the runtime complexity of \sysname to demonstrate that it incurs almost no overhead compared to heuristics. 

\paragraph{Inference. }
All cache operations, i.e., read, write (including eviction), are $O(1)$. 
\sysname adds two components. 
On the data path, it needs to collect features. However, many of the features it collects, such as the number of requests and hits, are already tracked in most production caches. 
The primary additional overhead introduced by \sysname is the hit-position distribution histogram. This histogram is maintained as an array of counters, with exactly one counter incremented per access, yielding a strict $O(1)$ time complexity. In \autoref{sec:eval:throughput}, we show that the feature collection has a negligible impact on throughput. In deployments that are very sensitive to computational overhead, the feature collection can use sampled requests. 
On the control path, \sysname performs inference, which has a time complexity of $O(T \cdot D)$, where $T$ is the number of trees (20 in our model) and $D$ is the maximum depth (9 in our model). Since the inference happens rarely and asynchronously, and each inference call takes less than 2 ms, there is no impact on request serving.

\paragraph{Training.} \rev{
The training cost of \sysname is entirely offline and consists of two stages: label generation and model fitting. Label generation is the dominant cost. For each training trace and cache-size ratio, we run S4-FIFO over a fixed grid of candidate configurations and select the configuration with the lowest miss ratio as the label. Given $N$ training traces, $R$ cache-size ratios, $G$ candidate configurations, and $L$ requests per trace, label generation costs $O(N \cdot R \cdot G \cdot L)$. In our evaluation, $G$ is a small constant determined by the parameter grid. Moreover, the grid search is embarrassingly parallel across traces, cache sizes, and configurations. Model fitting is substantially cheaper. Once the labels have been generated, we train a small GBDT using one compact feature vector per trace/cache-size pair. The model contains 20 trees with a maximum depth of 9 and predicts over 18 representative parameter sets. Thus, retraining is a one-time offline cost and does not affect the request-serving path. Deployments can directly use the pre-trained model. Retraining is only needed when operators want to incorporate additional training data. \revv{Further details can be found in Appendix \autoref{appx:appendix-pretrain}.}
}

\paragraph{Storage. } 
Unlike object-level learning, which requires maintaining feature vectors per object, \sysname uses only 73 global features in total. Beyond the 2-bit counter inherited from S3-FIFO, no additional per-object metadata is needed. 
Including model parameters, the total storage overhead introduced by learning is on the order of only tens of kilobytes. 
The dominant storage overhead in \sysname arises from the ghost queue. Each ghost entry requires 8 bytes to record the object ID (or fingerprint) and its insertion timestamp. For a cache holding 1 million objects, the ghost queue can therefore consume several megabytes of memory. Nonetheless, as long as objects are relatively large (e.g., larger than 200 bytes), this overhead remains small compared to the total data volume.

\section{Evaluation}

In this section, we use \sysname as a case study to answer the following questions:
\begin{itemize}
    \item Do augmented heuristics expose sufficient headroom for efficiency improvement?
    \item Can we accurately learn the optimal parameters of augmented heuristics to realize these efficiency gains?
    \item Do learning-augmented heuristics improve robustness?
    \item Do learning-augmented heuristics reduce throughput?
    \item Is a single pre-trained model sufficient across workloads, and why?
\end{itemize}

\subsection{Methodology}
\paragraph{Workloads.}
We evaluate \sysname using 5,175 production traces collected from 14 sources (\autoref{tab:dataset}). We exclude very small traces with fewer than 100,000 objects, as these typically correspond to low-priority workloads (e.g., development traffic). Notably, many learned caches, such as LRB and 3L-Cache, perform extremely poorly on these short traces, which skews both average efficiency and robustness metrics. In contrast, S4-FIFO is insensitive to trace length because it relies on aggregated cache-level features. Consequently, including all traces would make S4-FIFO’s results even stronger. 

\begin{table}[t]
\centering
\small
\caption{Summary of datasets used in our evaluation. We filtered out traces with fewer than 100,000 objects. \vspace{-0.8em}}
\label{tab:dataset}

\begin{tabular}{lrrr} 
\toprule
{Dataset} & {Approx time} & {Cache type} & {\# Traces} \\
\midrule
MSR~\cite{msr-dataset,msr-paper} & 2007 & Block & 14 \\
FIU~\cite{fiu} & 2008-11 & Block & 7 \\
CloudPhysics~\cite{shards} & 2015 & Block & 104 \\
Systor~\cite{systor-dataset,systor-paper} & 2017 & Block & 6 \\
CDN 1 & 2018 & Object & 163 \\
Tencent Photo~\cite{tencentPhoto-dataset,tencentPhoto-paper} & 2018 & Object & 2 \\
WikiMedia CDN~\cite{wikimedia-dataset} & 2016-19 & Object & 4 \\
Tencent CBS~\cite{tencentBlock-dataset,tencentBlock-paper} & 2020 & Block & 3370 \\ 
Alibaba~\cite{alibabaBlock-traces,alibabaBlock-paper1,alibabaBlock-paper2} & 2020 & Block & 552 \\ 
Twitter~\cite{DBLP:conf/osdi/YangYR20} & 2020 & KV & 50 \\
CDN 2 & 2021 & Object & 890 \\
Meta Storage~\cite{cachelib-workload} & 2022 & Block & 5 \\
Meta KV~\cite{cachelib-workload} & 2022 & KV & 5 \\
Meta CDN\cite{cachelib-workload} & 2023 & Object & 3 \\
\bottomrule
\end{tabular}
\end{table}

\begin{figure*}[t]
  \centering
    \begin{subfigure}[t]{0.48\linewidth}
    \centering
    \includegraphics[width=\linewidth]{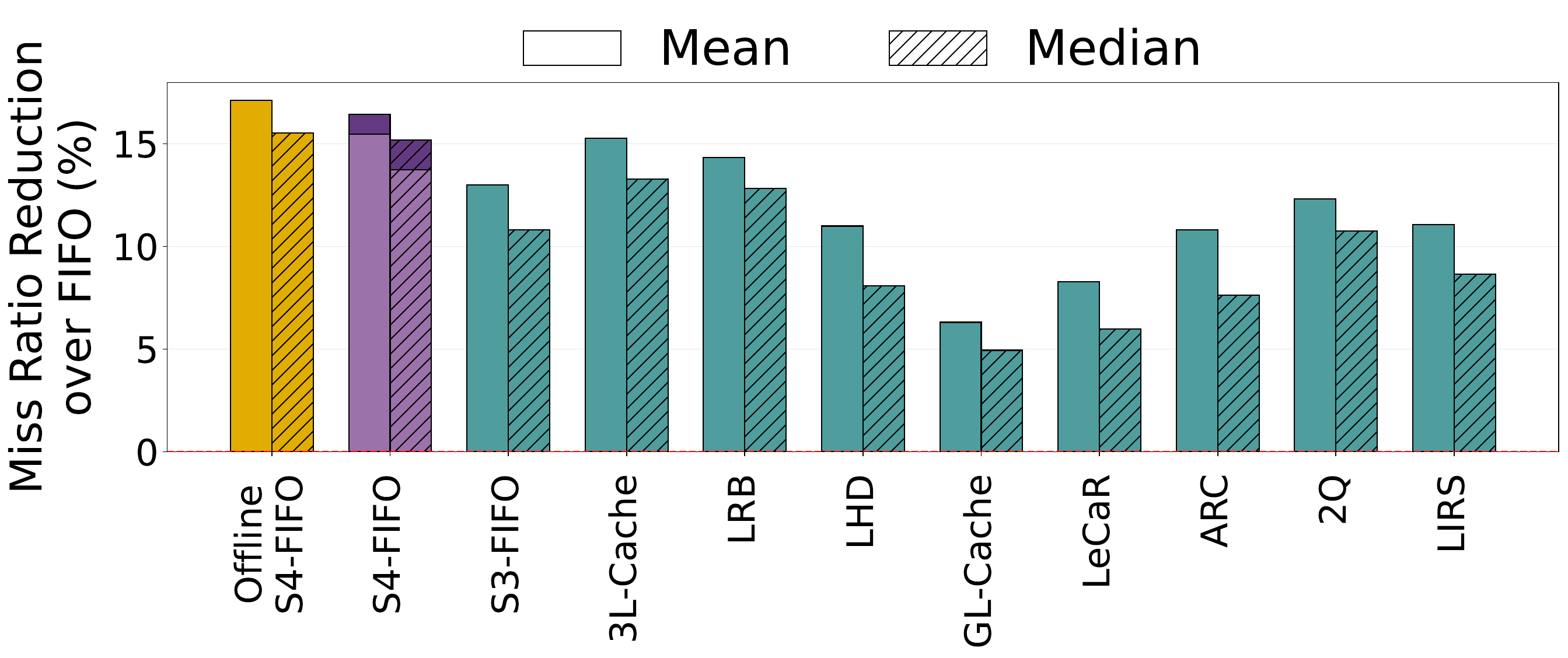}
    \caption{Large cache (10\% of working set)}
    \vspace{-0.8em}
  \end{subfigure}\hfill
  \begin{subfigure}[t]{0.48\linewidth}
    \centering
    \includegraphics[width=\linewidth]{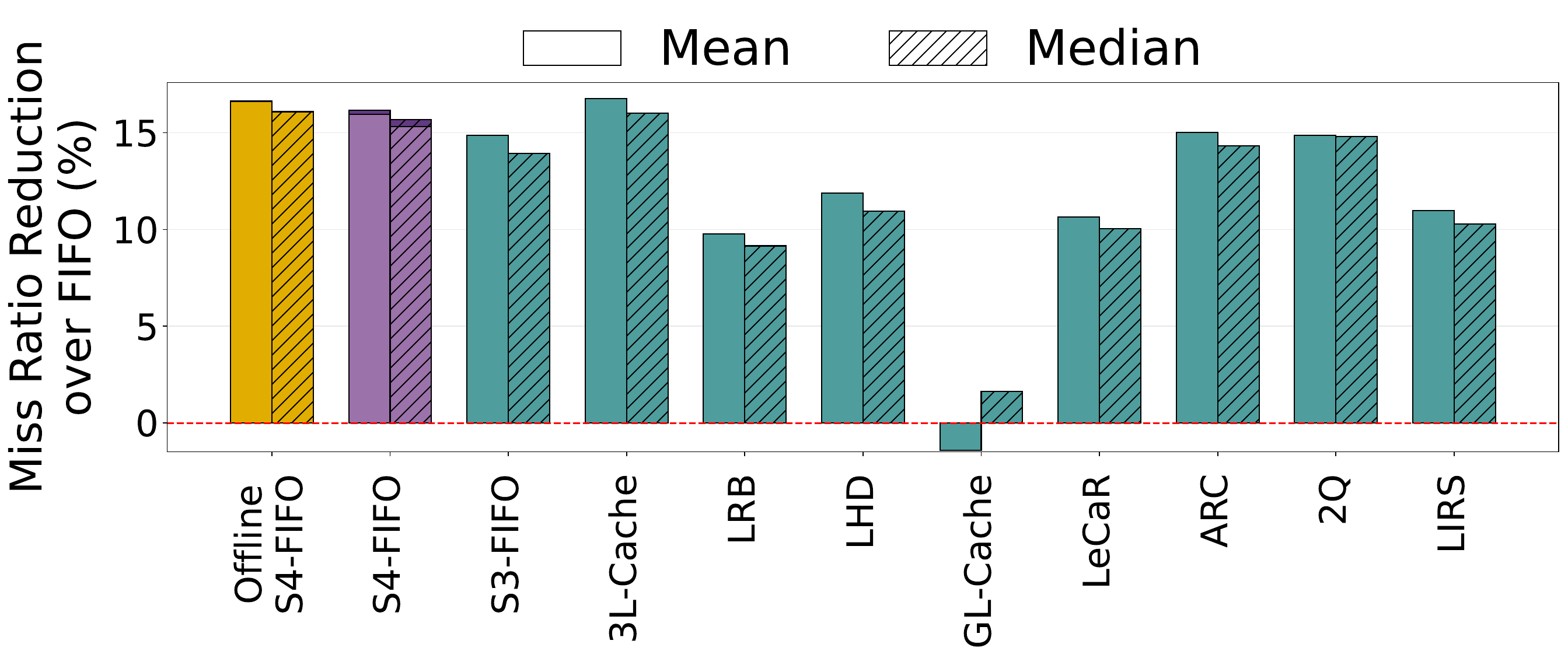}
    \caption{Small cache (0.1\% of working set)}
    \vspace{-0.8em}
  \end{subfigure}
  \caption{Mean and median miss ratio reduction from FIFO of different algorithms using the large and small cache sizes. \sysname is significantly better than all state-of-the-art algorithms at the large cache size, while on par with the best of all algorithms at the small cache size.}
  \vspace{-0.8em}
  \label{fig:efficiency}
\end{figure*}

\paragraph{Experiment setup. }
We randomly split the traces into two sets: 4140 for training and 1035 for evaluation. 
For each trace in the training set, we run \sysname with the default parameter combinations to collect training samples of (features, label) pairs. For each trace in the test set, we start the cache using \sysname with the default parameter combinations, run for 20\% of the trace to collect features, then make one prediction and apply the predicted parameters to the rest of the trace. To ensure a fair comparison with the offline optimum obtained via grid search, we restrict ourselves to one prediction per trace, since the offline optimum also uses a single parameter setting for the entire trace. 

We compare \sysname with S3-FIFO~\cite{s3fifo}, LIRS~\cite{LIRS}, ARC~\cite{ARC}, LRB~\cite{LRB}, 3L-Cache~\cite{3L-Cache}, LHD~\cite{LHD}, GL-Cache~\cite{GLcache}, and LeCaR~\cite{LeCaR}. Besides, we have also evaluated SIEVE~\cite{sieve}, GDSF~\cite{cherkasova_improving_1998}, and TinyLFU~\cite{TinyLFU}; however, their results are worse than the ones in the figure. Due to space limits and for ease of plotting, we omit them in the figure. We evaluate each algorithm at three cache sizes—0.1\%, 1\%, and 10\% of the trace’s working-set size—but, for space reasons, we only present results for the smallest (0.1\%) and largest (10\%) caches. The results for the medium size (1\%) consistently fall between these two. 

All miss ratio results are from libCacheSim because most state-of-the-art eviction algorithms are not available in open-source caches such as Cachelib. Because our Cachelib implementation uses the same model as the simulator, \sysname shows almost identical miss ratios in the simulator and prototype. 
We report throughput results by comparing \sysname with highly optimized LRU, 2Q, S3-FIFO, and TinyLFU using CacheBench from Meta. The experiments were performed on a 20-node cluster with 32 cores and 192GB of memory. All experiments are repeated three times, and we report the mean result. 


\paragraph{Metrics.}
We evaluate using the following metrics. \\
(i) \emph{Efficiency}: Following prior work~\cite{s3fifo}, we report the miss ratio reduction over FIFO, computed as $\frac{MR_{\text{FIFO}} - MR_{\text{algo}}}{MR_{\text{FIFO}}}$. We present the mean reduction in miss ratio across the 1,035 traces in the test set. \\
(ii) \emph{Robustness}: We report miss ratio reduction for the worst-case workload and for the $10^{\text{th}}$-percentile workload. \\
(iii) \emph{Overhead}: We use Cachelib throughput as our primary overhead metric.

\begin{figure*}[t]
  \centering
  \begin{subfigure}[t]{0.24\linewidth}
    \centering
    \includegraphics[width=\linewidth]{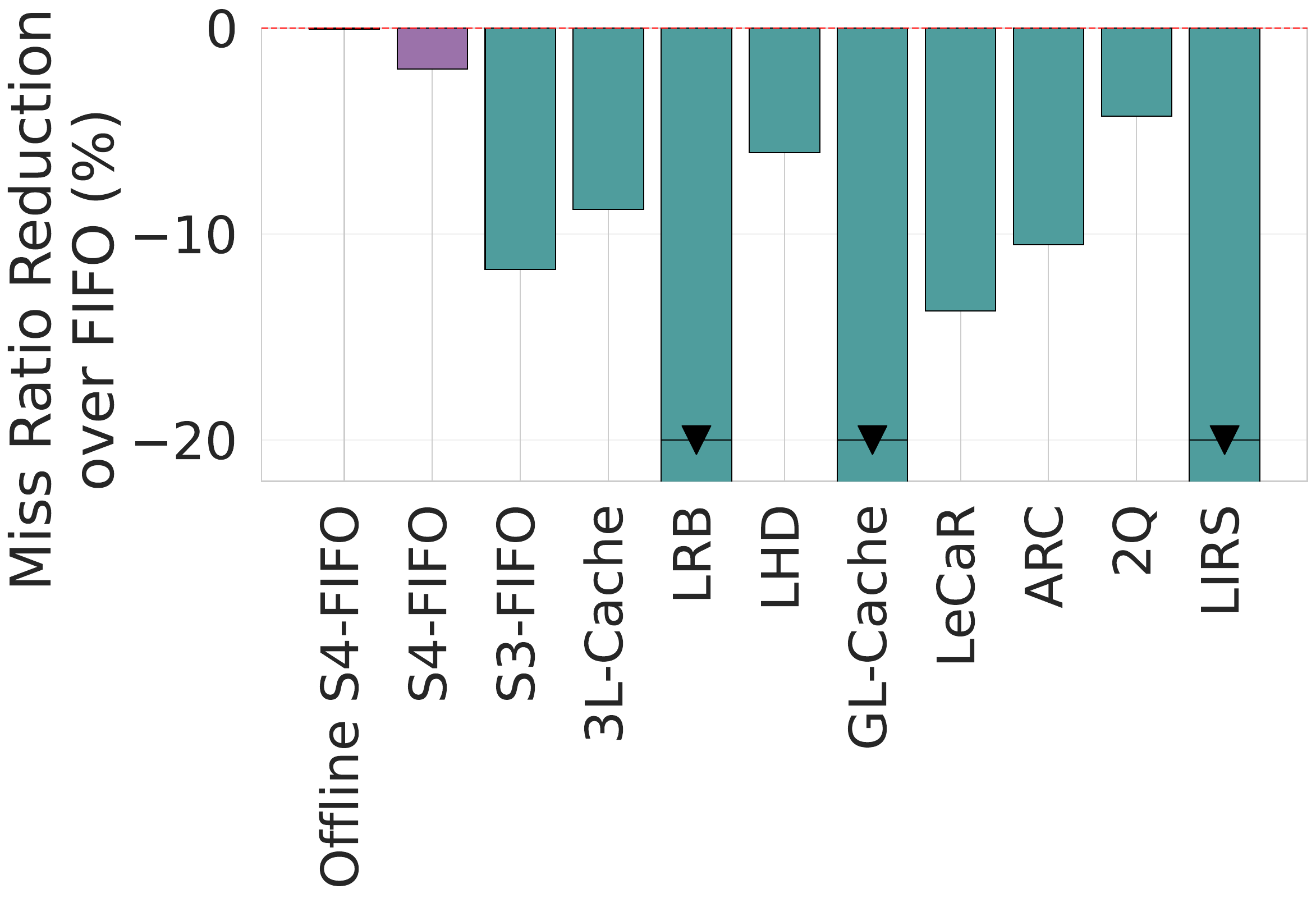}
    \caption{Worst case, large cache}
    \label{fig:robustness:worst_large}
    \vspace{-0.5em}
  \end{subfigure}\hfill
  \begin{subfigure}[t]{0.24\linewidth}
    \centering
    \includegraphics[width=\linewidth]{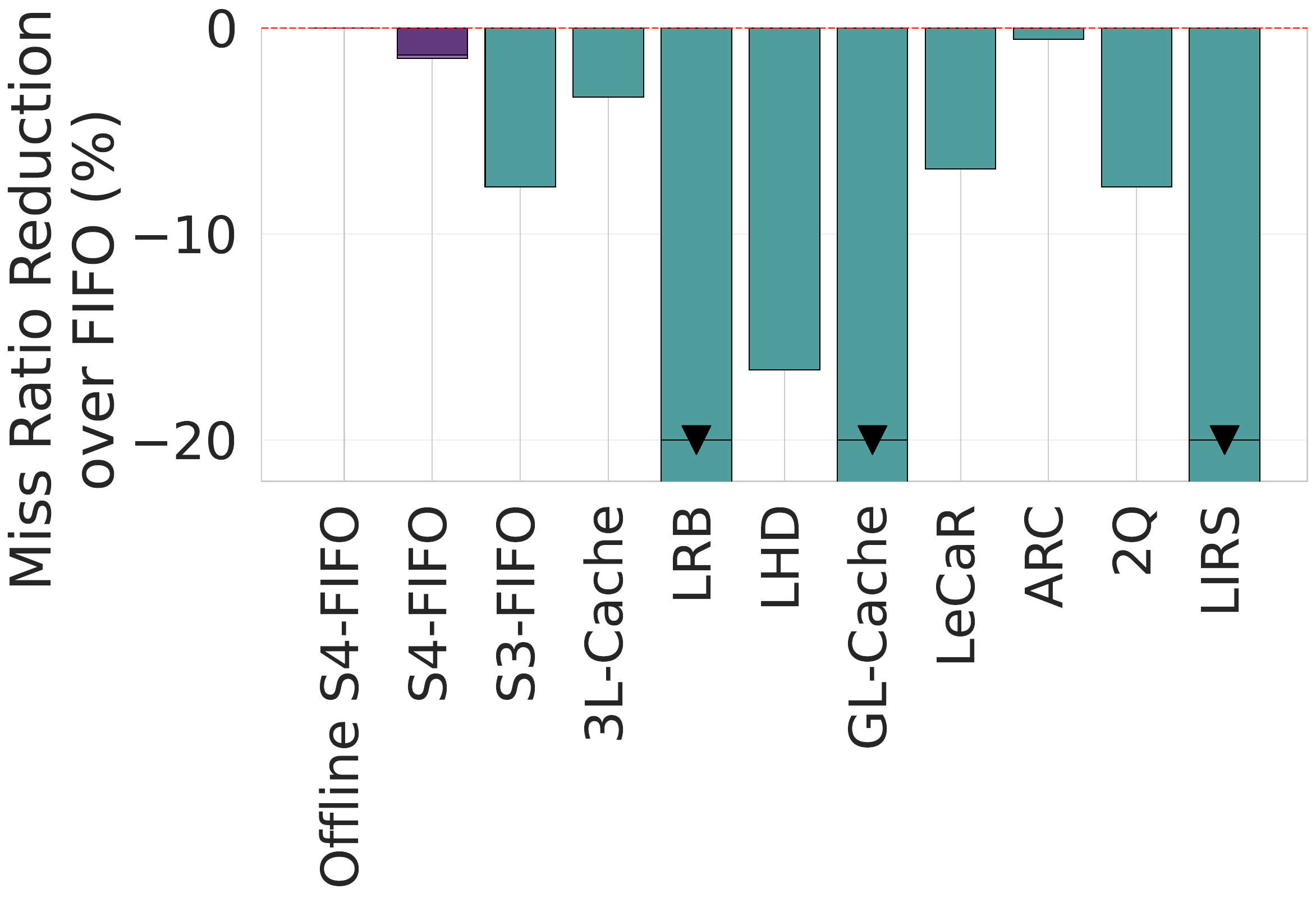}
    \caption{Worst case, small cache}
    \label{fig:robustness:worst_small}
        \vspace{-0.5em}
  \end{subfigure}\hfill
  \begin{subfigure}[t]{0.24\linewidth}
    \centering
    \includegraphics[width=\linewidth]{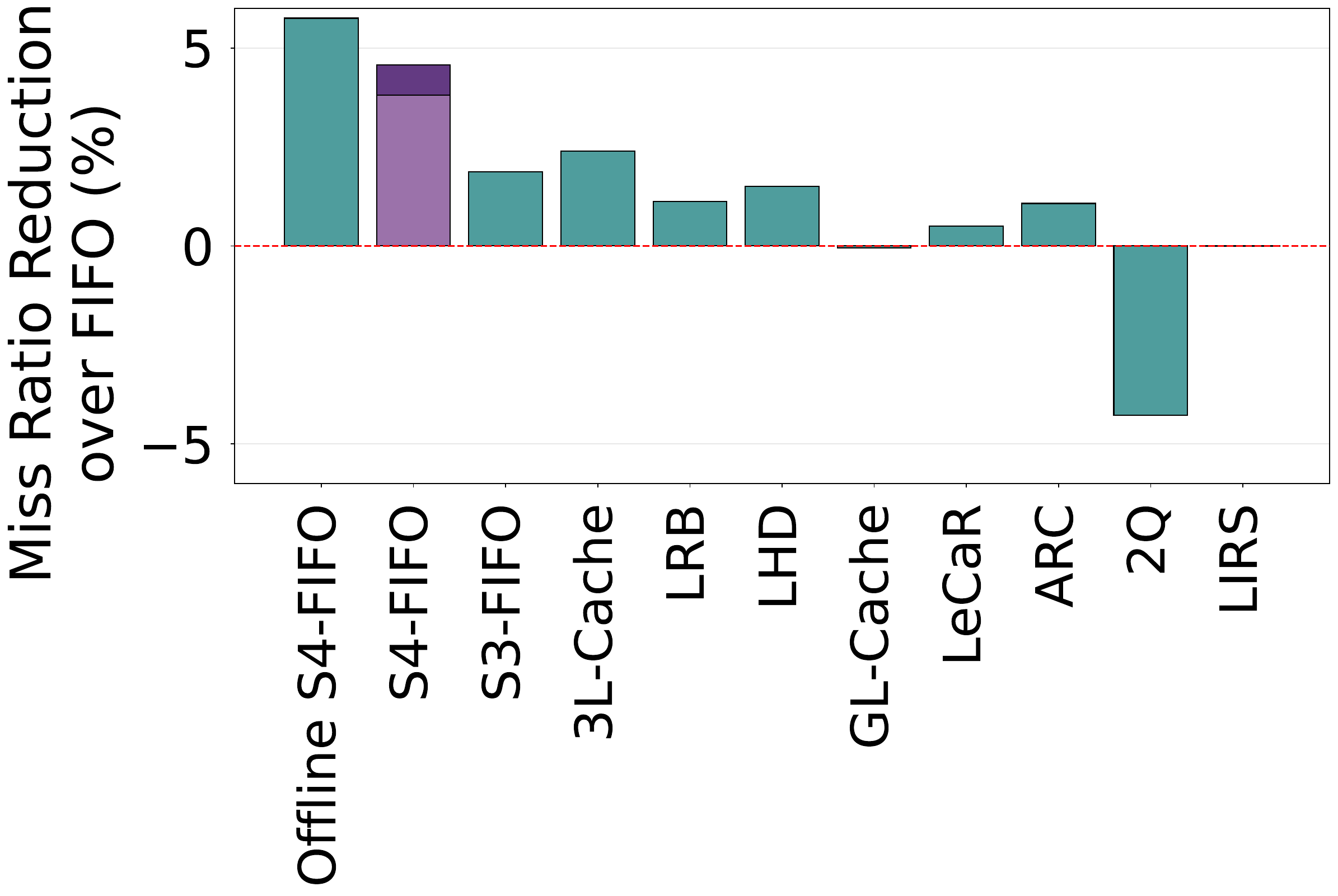}
    \caption{10$^{\text{th}}$-percentile, large cache}
    \label{fig:robustness:p10large}
        \vspace{-0.5em}
  \end{subfigure}\hfill
  \begin{subfigure}[t]{0.24\linewidth}
    \centering
    \includegraphics[width=\linewidth]{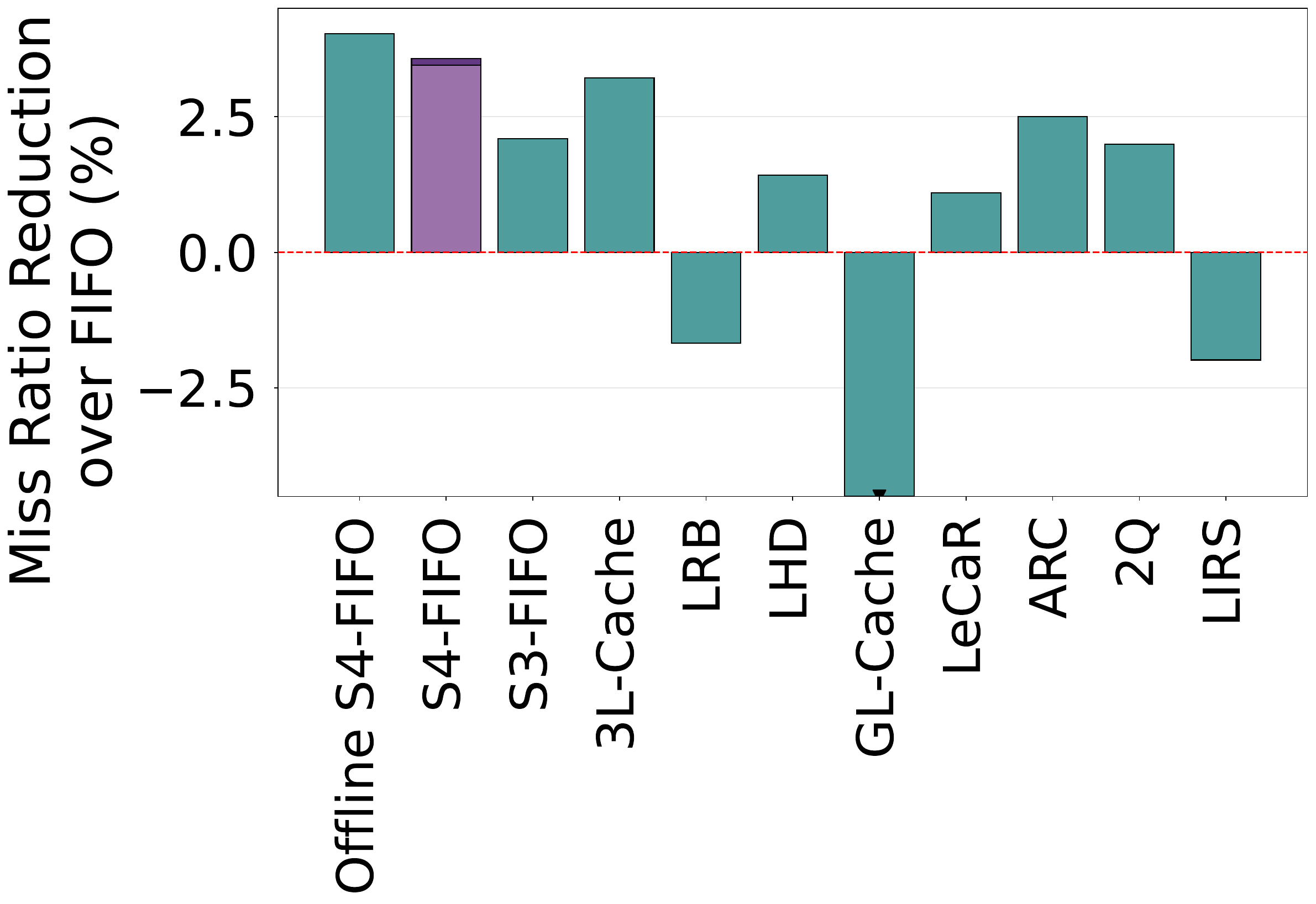}
    \caption{10$^{\text{th}}$-percentile, small cache}
    \label{fig:robustness:p10small}
        \vspace{-0.5em}
  \end{subfigure}
  \caption{Efficiency on the worst case and 10$^{\text{th}}$-percentile trace. A negative value indicates the algorithm has a miss ratio higher than FIFO. \sysname is more robust than all state-of-the-art algorithms. }
  \vspace{-1.5em}
  \label{fig:robustness}
\end{figure*}

\subsection{Efficiency}

Efficiency is the primary performance metric for cache eviction. \autoref{fig:efficiency} reports the mean and median miss-ratio reduction over FIFO for both small and large caches. Since \sysname uses the first 20\% of each trace to collect features before making a prediction, we evaluate two variants. In the online variant (v1, light purple), the first 20\% of requests are served using the default parameters, and the remaining 80\% are served using the predicted parameters. In the retrospective variant (v2, dark purple), the predicted parameters are applied to the entire trace; this isolates the quality of the learned configuration from the cost of the observation window. Across both cache sizes, v2 consistently outperforms v1, indicating that the predicted parameters improve efficiency and that the observation window accounts for the gap between the two variants.

\paragraph{Offline \sysname.} 
Unlike many learned caches, such as 3L-Cache, LRB, and LHD, which use sampling to select eviction candidates and can evict any object in the cache, learning-augmented heuristics like \sysname are constrained by their underlying data structures: objects must pass through at least one queue before they can be evicted. Nevertheless, \autoref{fig:efficiency} shows that offline \sysname, which selects parameters via grid search, consistently achieves the highest efficiency. This suggests that the data structures used in \sysname do not prevent it from attaining high efficiency. 
When comparing \sysname with offline \sysname, we find that the second variant, which runs the full trace using predicted parameters (same as offline \sysname), achieves very similar miss ratio reduction as offline \sysname with at most 0.2\% difference in mean and median miss ratio reduction. This demonstrates the effectiveness of learning in \sysname. In \autoref{sec:eval:learning}, we will show that \sysname learns the best parameters most of the time, and even when it does not, the learned parameters are close to the best.

\paragraph{Comparison with state-of-the-art algorithms.}
\sysname is built on top of S3-FIFO; compared to S3-FIFO, it achieves a 26\% higher miss ratio reduction at the large cache size and an 8\% higher reduction at the small cache size. 
Across all state-of-the-art algorithms, 3L-Cache consistently delivers the best performance. Relative to 3L-Cache, \sysname attains slightly higher or comparable miss ratio reductions. For instance, at the large cache size, \sysname achieves an 8\% higher reduction. 
Compared with ARC, \sysname nearly doubles ARC’s improvement over FIFO at the large cache size. 
Relative to the large-cache setting, S4-FIFO’s performance at the small cache size is less impressive: it is slightly worse than 3L-Cache, though still substantially better than the other state-of-the-art algorithms. 
For instance, \sysname improves LRB’s mean reduction ratio from 9.8\% to 16.2\%. 
The slight advantage of 3L-Cache at small cache sizes stems from the fact that the cached objects are all highly popular, providing rich signals for object-level learning. \revv{However, object-level learning incurs high overhead and low throughput. In our simulator, 3L-Cache is 17.3$\times$ slower than \sysname on average (max 274$\times$)}. 
In summary, \sysname achieves a high efficiency via its superior learning methods and the pre-trained model.

\subsection{Robustness}
Historically, work on designing new cache eviction algorithms has focused on reducing miss ratio on a small set of traces or improving average performance. Far less attention has been given to the robustness of these algorithms. Yet robustness is crucial, because an algorithm can severely degrade performance when the workload is out of distribution relative to its design assumptions. 
In this section, we evaluate the robustness of state-of-the-art algorithms using the worst trace and $10^{\text{th}}$-percentile trace. Note that we do not consider the competitive ratio because the arbitrary workload used in competitive-ratio analysis rarely occurs in the real world, so we focus on the worst-case production trace. 

\autoref{fig:robustness:worst_large} and \autoref{fig:robustness:worst_small} show the miss ratio change relative to FIFO on the worst trace for each algorithm. We observe that all state-of-the-art eviction algorithms incur a higher miss ratio than FIFO on their worst-case trace. In particular, LRB and LIRS perform substantially worse than the others, increasing FIFO’s miss ratio by 20\% to 72\%. In contrast, \sysname increases FIFO’s miss ratio by only 0.8\% and 0.2\% at large and small cache sizes, respectively. For comparison, the next-best algorithm at the large cache size is 2Q, which raises FIFO’s miss ratio by 4.3\%.

\begin{figure}[t]
  \centering
    \begin{subfigure}[t]{0.48\linewidth}
    \centering
    \includegraphics[width=\linewidth]{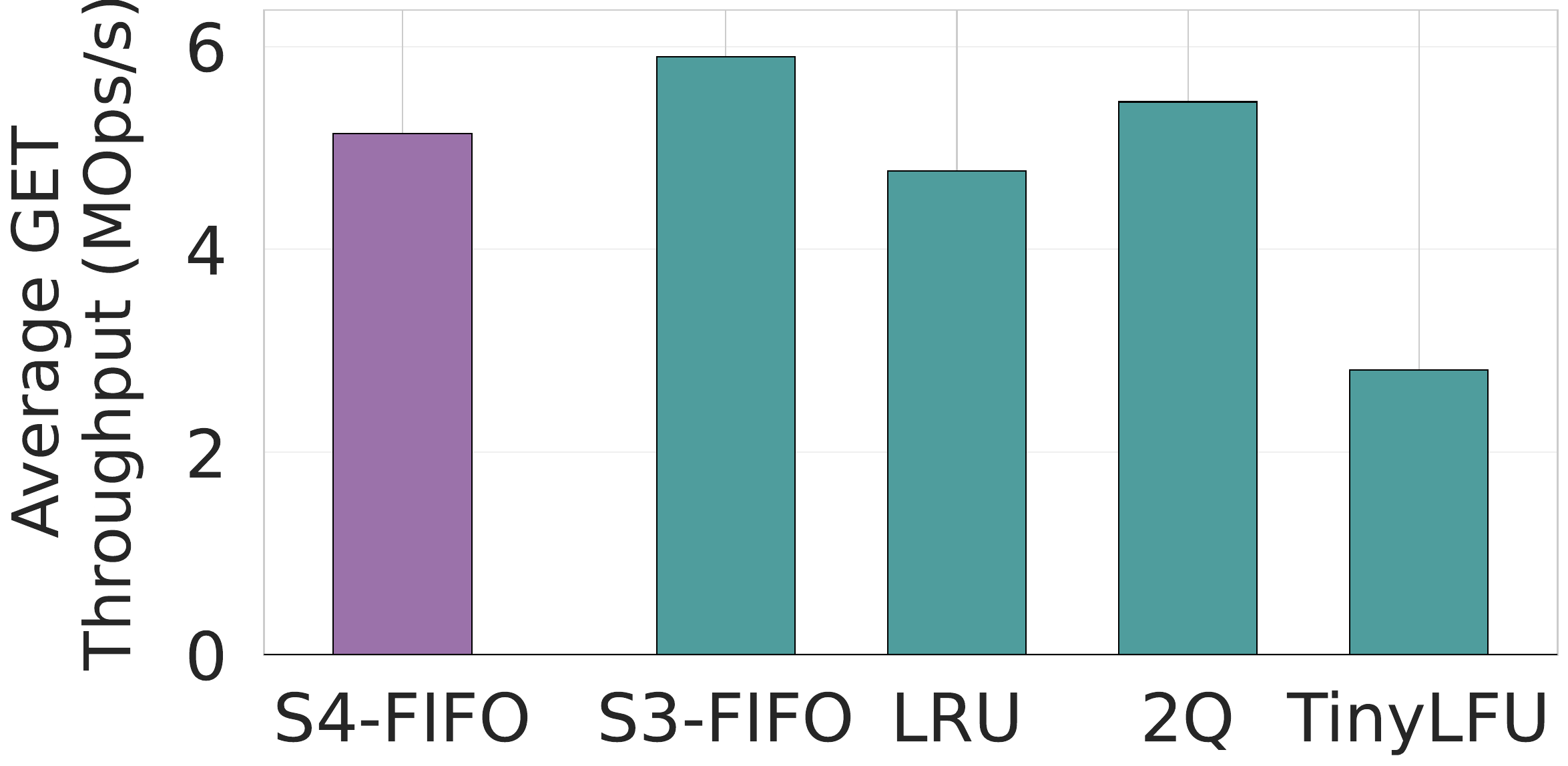}
    \vspace{-1em}
  \end{subfigure}\hfill
  \begin{subfigure}[t]{0.48\linewidth}
    \centering
    \includegraphics[width=\linewidth]{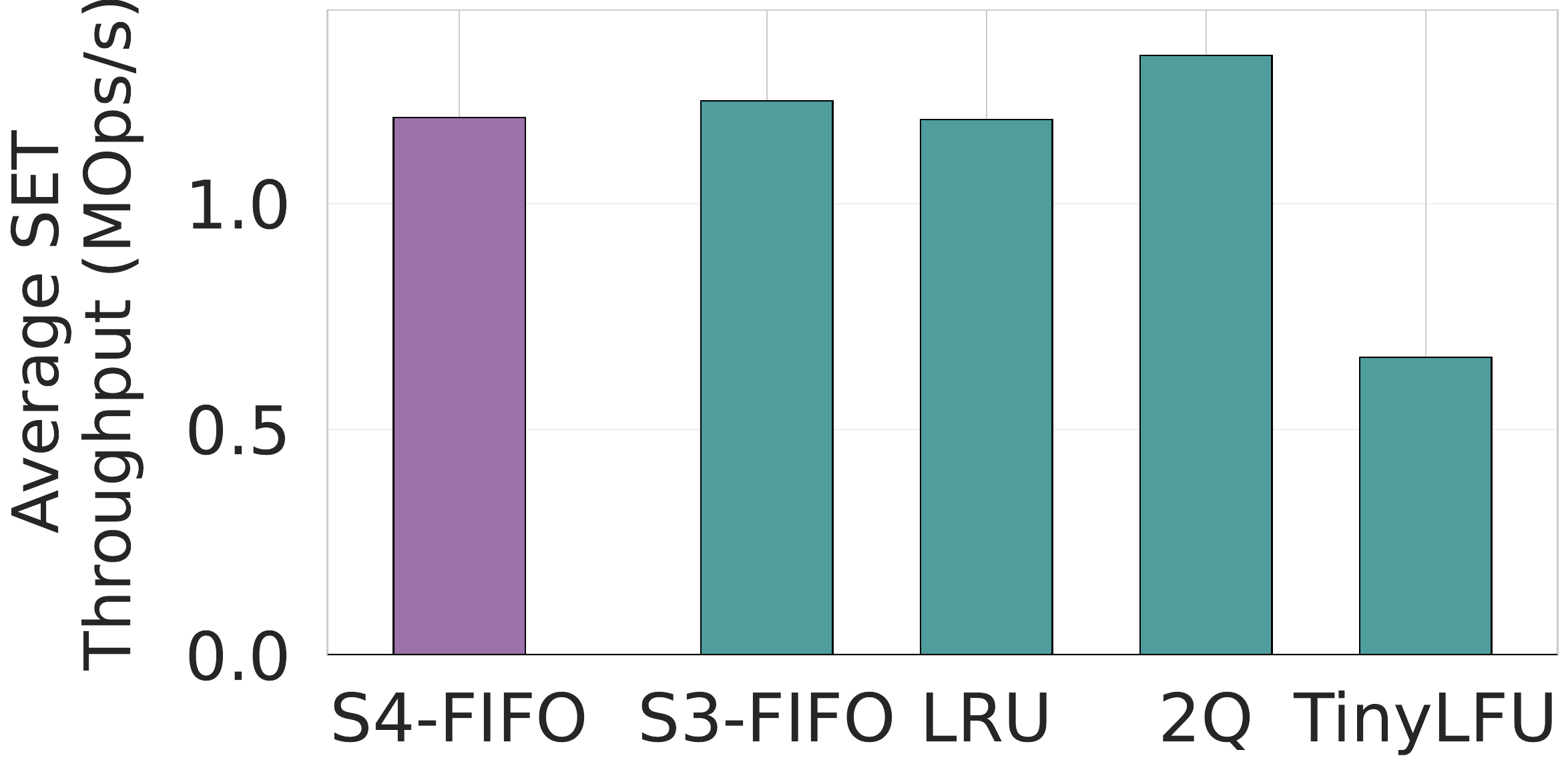}
    \vspace{-1em}
  \end{subfigure}
     \vspace{-0.5em}
  \caption{Set and Get throughput evaluated in Cachelib.}
   \vspace{-1.0em}
  \label{fig:throughput}
\end{figure}

\begin{figure*}[t]
  \centering
  \begin{minipage}[t]{0.32\linewidth}
    \centering
    \includegraphics[width=0.96\linewidth]{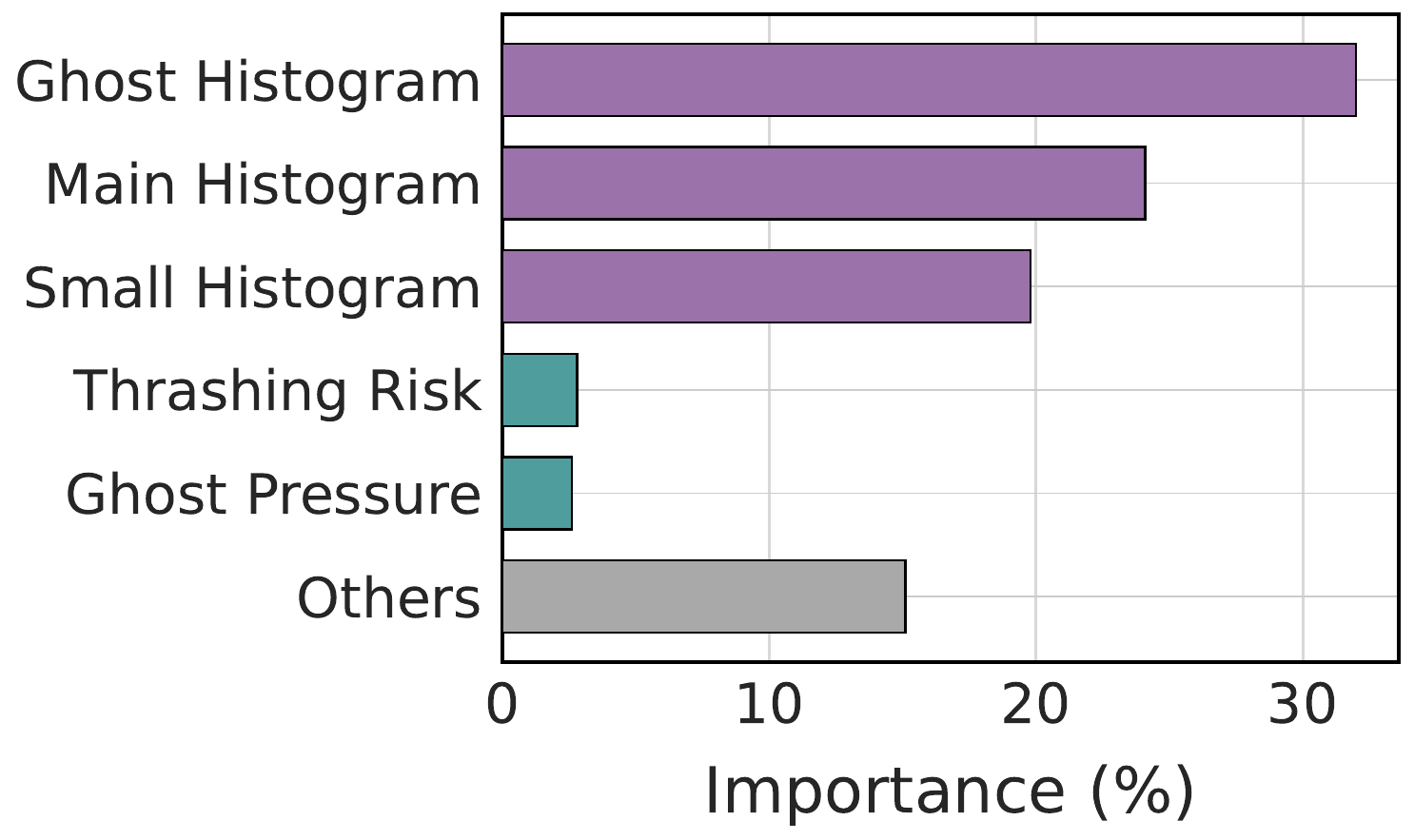}
    \caption{Feature importance by category.}
    \label{fig:why:featureimportance}
  \end{minipage}
  \begin{minipage}[t]{0.32\linewidth}
    \centering
    \includegraphics[width=0.96\linewidth]{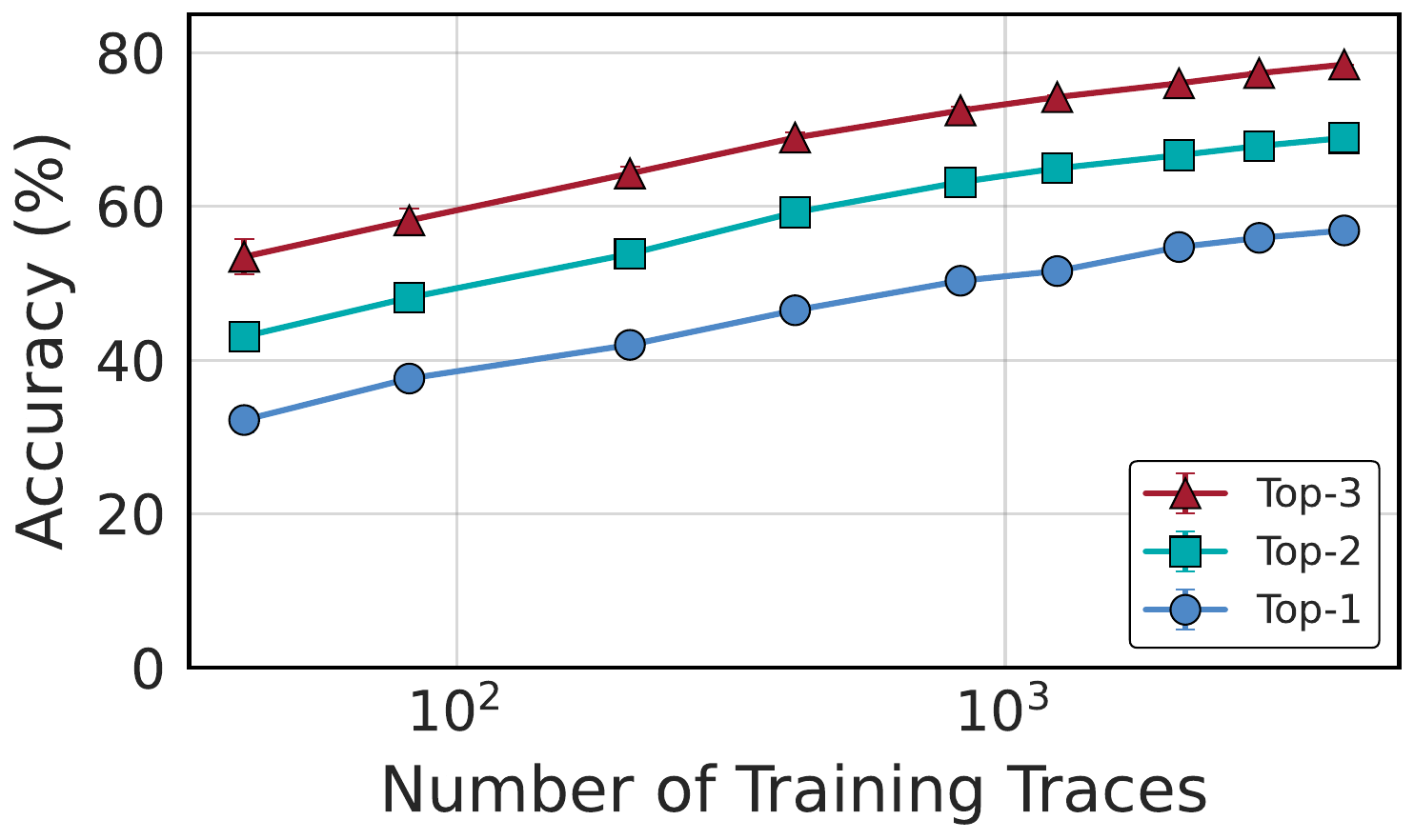}
    \caption{Impact of training data size.}
    \label{fig:learning:trainingsize}
  \end{minipage}
  \begin{minipage}[t]{0.32\linewidth}
    \centering
    \includegraphics[width=0.96\linewidth]{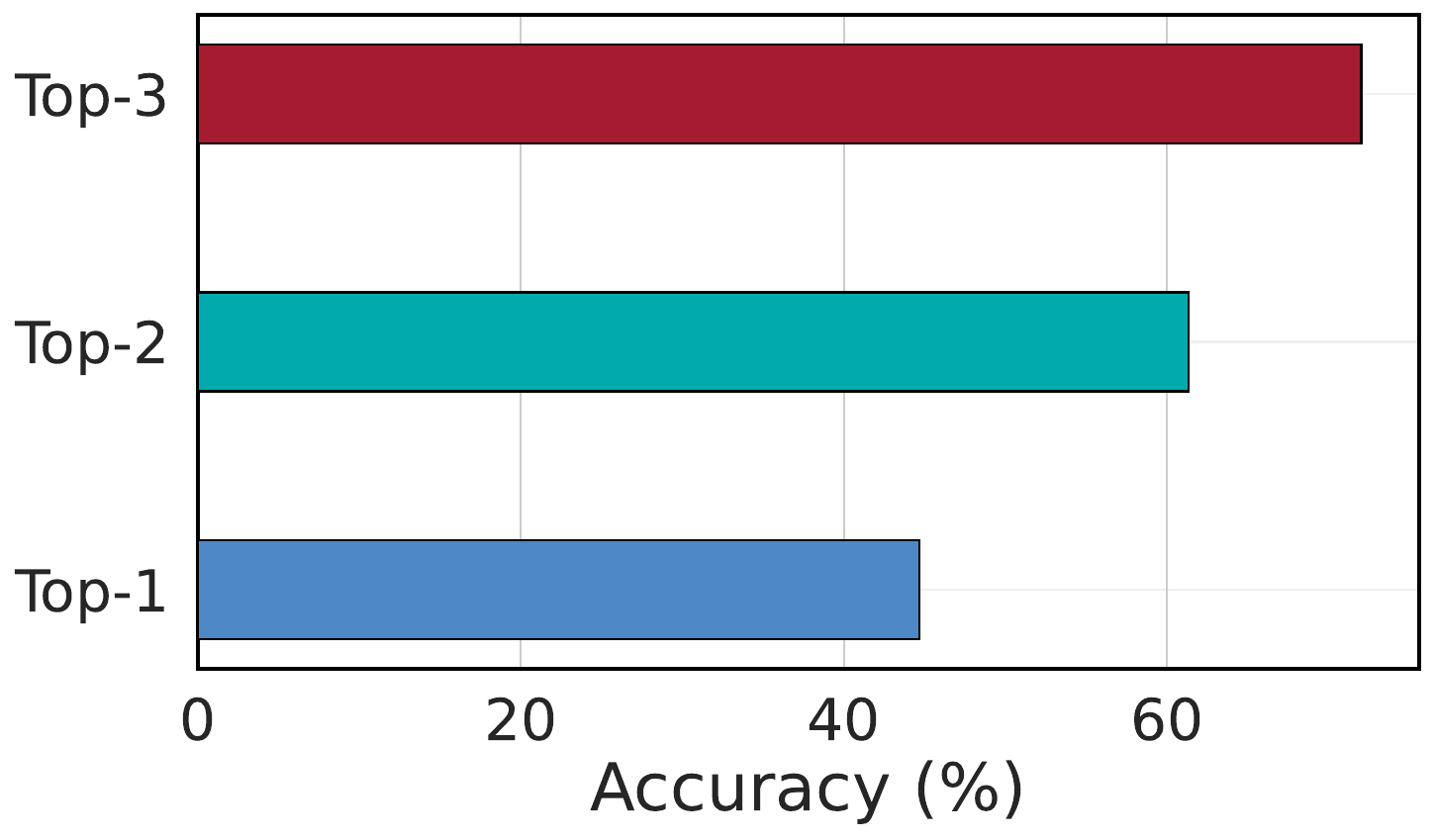}
    \caption{Prediction accuracy on the Twitter dataset on model pretrained on CDN2.}
    \label{fig:learning:generalization}
  \end{minipage}
\end{figure*}

Although the single worst-case trace may be unrepresentative, \autoref{fig:robustness:p10large} and \autoref{fig:robustness:p10small} show results on the 10th-percentile trace. On this metric, \sysname widens its lead over state-of-the-art algorithms. While some schemes, such as LRB and LIRS, still exhibit miss ratios higher than FIFO, \sysname reduces FIFO’s miss ratio by 4.2\% and 3.6\% at large and small cache sizes, respectively.

The robustness of \sysname arises from three core design choices. First, \sysname employs static FIFO queues and therefore avoids the pathological behaviors that can arise in adaptive algorithms~\cite{s3fifo}. Second, \sysname leverages learning to select its parameters, enabling it to sidestep the adversarial workloads that often degrade the performance of fixed, heuristic-based schemes. Third, \sysname's learning objective uses FIFO as an anchor to directly optimize for robustness. 


\vspace{-0.75em}
\subsection{Throughput}
\label{sec:eval:throughput}


\begin{figure}[t]
  \centering
    \begin{subfigure}[t]{0.48\linewidth}
    \centering
    \includegraphics[width=\linewidth]{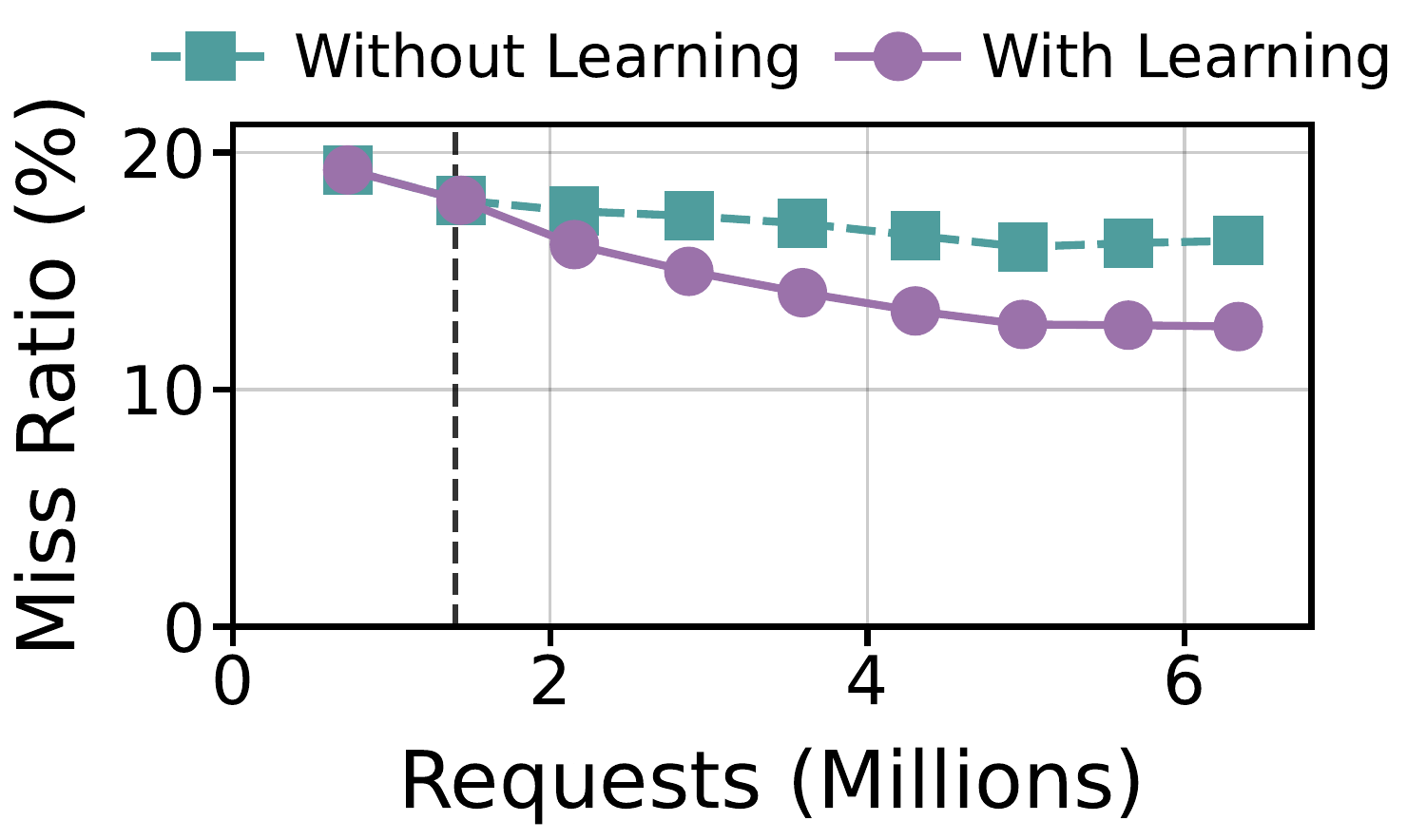}
    \caption{Cumulative miss ratio}
    \label{fig:learning:missratio:cumulative}
  \end{subfigure}\hfill
  \begin{subfigure}[t]{0.48\linewidth}
    \centering
    \includegraphics[width=\linewidth]{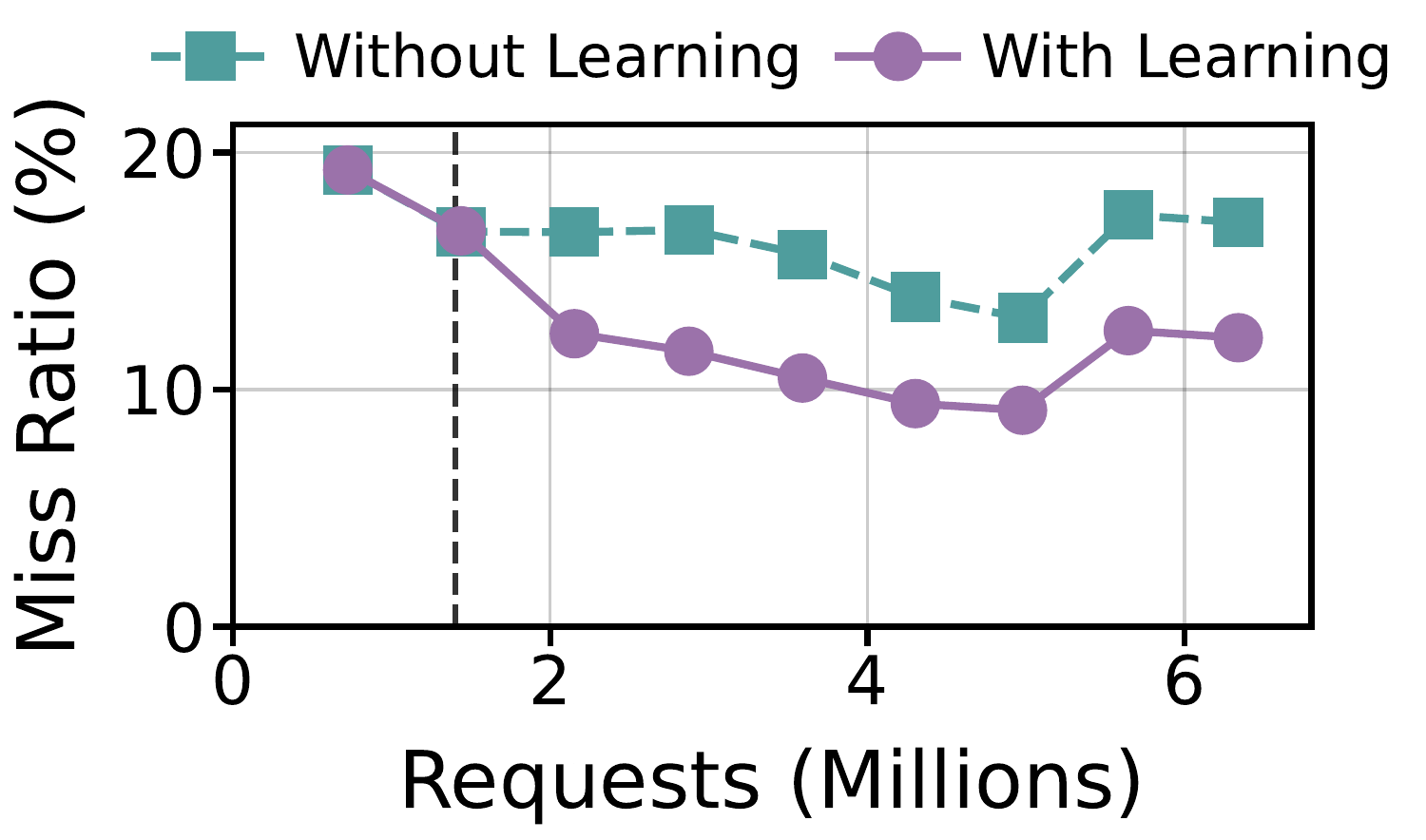}
    \caption{Interval miss ratio}
    \label{fig:learning:missratio:interval}
  \end{subfigure}
  \vspace{-0.75em}
  \caption{Miss ratio over time where the learned parameters significantly reduce the miss ratio. 
  }
  \vspace{-1.0em}
  \label{fig:learning:missratio}
\end{figure}

To assess whether S4-FIFO preserves the high throughput of lightweight heuristic policies, we measure its throughput using CacheBench and compare it against highly optimized Cachelib implementations of S3-FIFO, LRU, LRU2Q (a modified 2Q without a ghost queue), and TinyLFU. In this experiment, S4-FIFO collects features continuously for every request throughout the benchmark. This deliberately conservative configuration provides an upper bound on the overhead of feature collection. \revv{
In practice, periodic or sampled feature collection could further reduce this overhead; exploring these optimizations is left to future work.} We use CacheBench's CDN workload and four Graph Cache workloads, run each benchmark with 48 threads, and report the average throughput across all five workloads in \autoref{fig:throughput}.

Even under this deliberately heavy configuration, S4-FIFO sustains throughput comparable to conventional heuristics, with only a modest slowdown from continuous feature extraction. Achieving heuristic-level throughput is challenging for typical learned caches, which often require critical-path inference or online training. S4-FIFO avoids these costs by collecting features outside the critical section and performing inference infrequently and asynchronously using an offline-trained model. As a result, its critical path remains lightweight, preserving throughput and simplifying deployment. By contrast, TinyLFU incurs additional overhead because its count-min sketch must be updated within the critical section to ensure correctness, resulting in significant throughput degradation.

\subsection{Why does Learning Work?}
\label{sec:eval:learning}

\vspace{-0.5em}
As a simple algorithm with static queue sizes, \sysname achieves both higher efficiency and robustness compared to state-of-the-art algorithms. This section deep dives into how learning helps \sysname achieve this and why one pre-trained model is sufficient. 

\vspace{-1em}
\paragraph{Miss ratio over time. } 
\autoref{fig:learning:missratio} illustrates the effect of how learned parameters affect miss ratio over time for a representative trace. 
The vertical dashed line marks the point at which \sysname switches to the predicted parameters, which shrink the small-queue size to 5\% of the cache and skip frequency-counter increments for 25\% of the items at the front of the small queue. 
After this parameter change, \sysname (solid line) quickly diverges from the baseline (dashed line): the interval miss ratio drops substantially and remains consistently lower for the remainder of the trace, and the cumulative miss ratio curve steadily widens its gap relative to the baseline. This example shows that although parameters remain static (i.e., not updated per-miss), learning a set of parameters for each workload can deliver a substantial reduction in miss ratio compared to using a fixed set of parameters.

\begin{figure}[t]
  \centering
  \vspace{-1.0em}
  \begin{subfigure}[t]{0.48\linewidth}
    \centering
    \includegraphics[width=\linewidth]{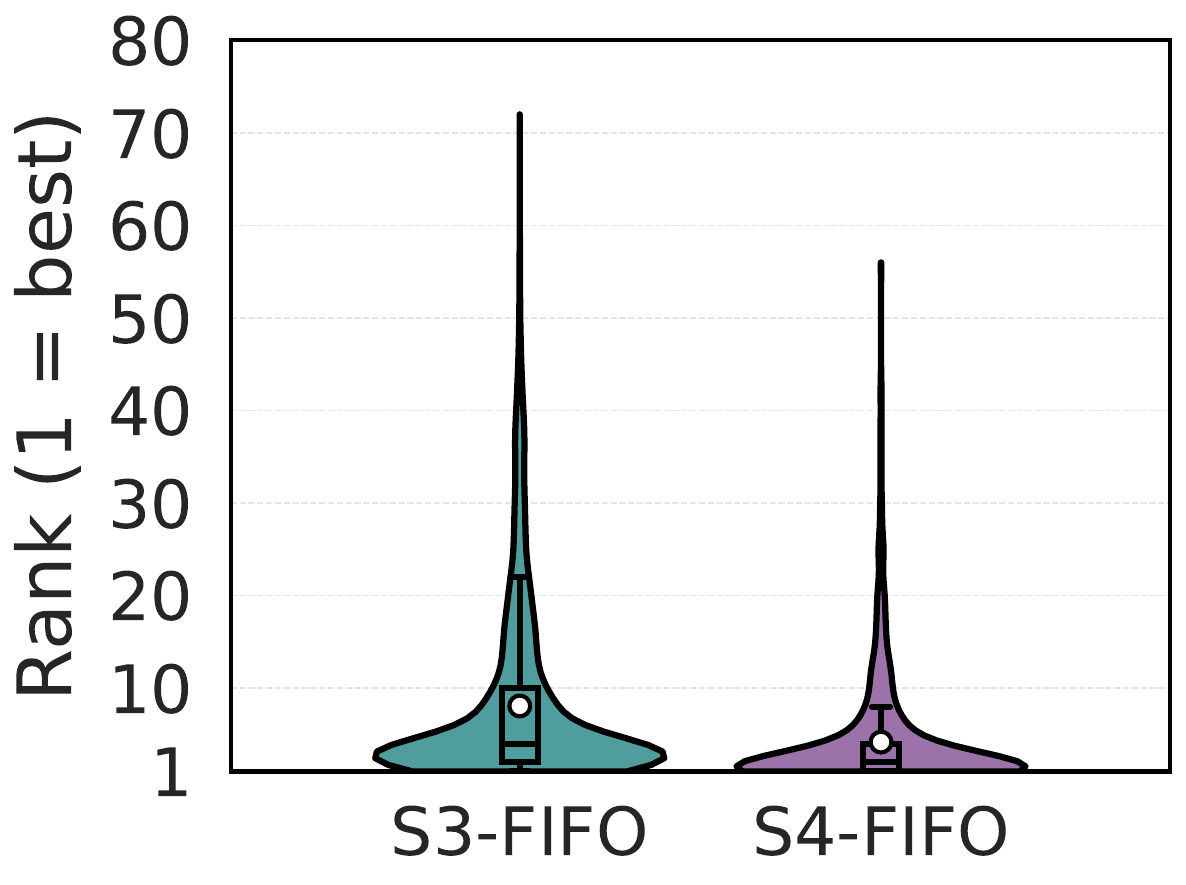}
    \caption{Large cache}
    \label{fig:learning:rank:large}
  \end{subfigure}
    \begin{subfigure}[t]{0.48\linewidth}
    \centering
    \includegraphics[width=\linewidth]{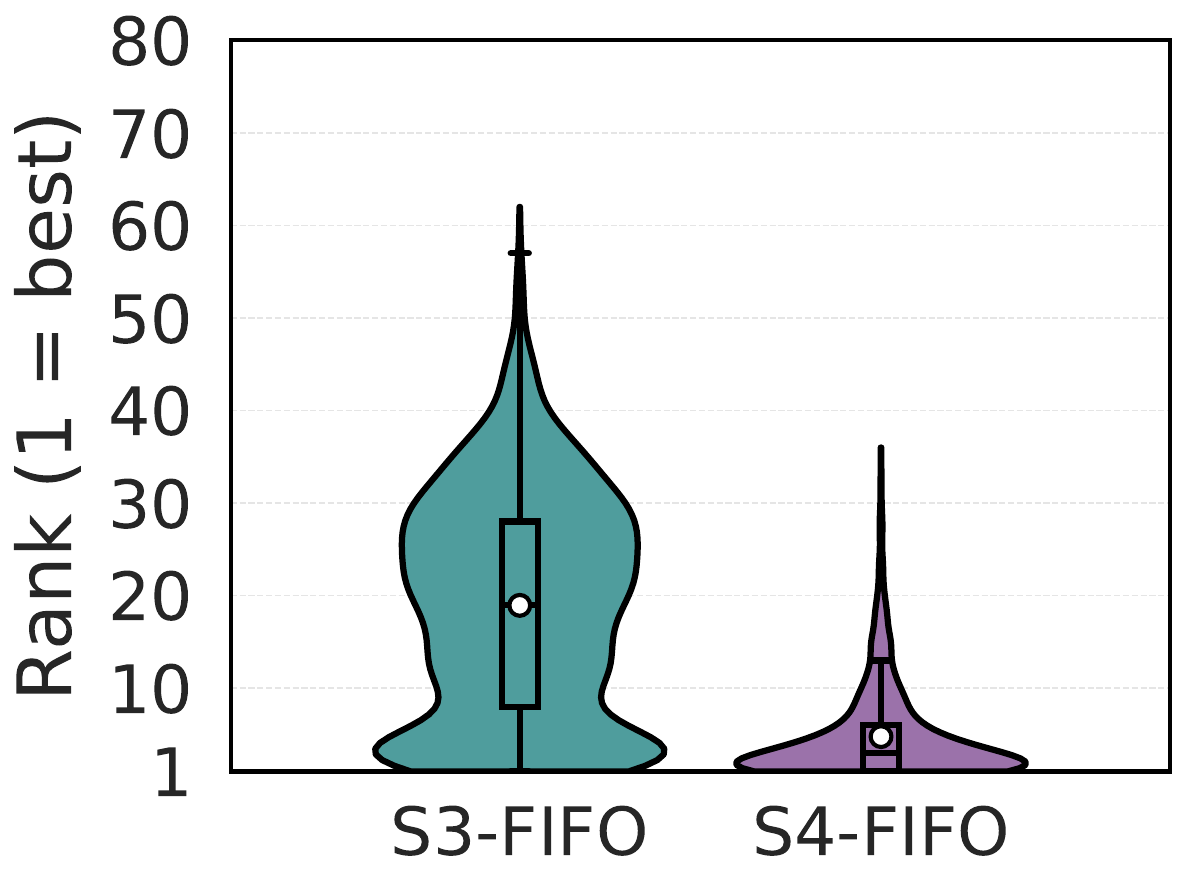}
    \caption{Small cache}
    \label{fig:learning:rank:small}
  \end{subfigure}\hfill
  \caption{Rank distribution of the default (S3-FIFO) and predicted parameters (S4-FIFO) from optimal grid search configuration.}
  \vspace{-1.5em}
  \label{fig:learning:rank}
\end{figure}

\paragraph{Rank of predicted parameters. }
\autoref{fig:learning:rank} compares the grid-search rank distributions of the default \sysname parameters (most are inherited from S3-FIFO) with the learned parameters, where a lower rank indicates better performance (rank 1 is the best). In both small and large cache sizes, the predicted parameters from \sysname are consistently close to the best configuration across traces. In contrast, the default parameters have a substantially wider distribution that extends to much worse ranks, indicating that using the same parameters across traces leaves significant headroom for improvement.

\paragraph{Feature importance. } We compute the feature-importance scores from the trained decision tree. As shown in \autoref{fig:why:featureimportance}, the histograms of all three queues (ghost, main, and small) are the most important. Together, these histogram features account for ~75\% of the total importance, indicating that the shape of hit distributions within each queue (workload locality) is the strongest signal for predicting good configurations. The remaining composite features, such as thrashing risk (working set size estimate) and ghost pressure (hits found on the ghost), account for ~25\% of the importance.

\paragraph{Training data size. }
To understand why a single pre-trained model works, we measure how prediction accuracy changes with the number of training traces. For each training dataset size, we pre-train a new model and evaluate on the same test dataset. 
We measure top-1, top-2, and top-3 accuracy, which are the fractions of traces in which the predicted parameters are ranked top-1, top-2, or top-3, respectively. 
\autoref{fig:learning:trainingsize} shows that all three metrics monotonically increase with more training data. With several thousand traces, top-1 approaches 60\%, top-2 approaches 70\%, and top-3 approaches 80\%. Although performance starts to plateau, increasing the training dataset size further is likely to make the pre-trained model increasingly accurate.

\paragraph{Cross-dataset generalization. }
The training and test traces we have been using so far are generated using a random split, without considering the dataset (source). 
This is acceptable because each dataset contains many diverse traces, and traces within a single dataset can be more diverse than those across datasets.  
To verify whether pre-trained models can generalize across datasets, we train a model on the CDN2 dataset (object cache) and evaluate on the Twitter dataset (key-value cache). 
\autoref{fig:learning:generalization} shows that the model pre-trained on one dataset can be directly used on a different one, and the knowledge learned is generalizable as long as the training dataset is large and diverse enough.

\paragraph{Generality across eviction algorithms.} \rev{
The principles of LAH can be applied to other policies that expose a small set
of configuration knobs. For example, CacheLib's LRU exposes configurable behaviors such as node-promotion frequency, whether metadata is updated on reads, writes, or contention, and the
insertion position of new objects in the LRU queue. 2Q~\cite{2Q} exposes the
relative sizes of its hot and cold queues, ARC~\cite{ARC} exposes the balance
between its recency and frequency regions, and LeCaR~\cite{LeCaR} exposes the
weights assigned to its eviction experts. Extending LAH to these policies
requires policy-specific feature engineering. For example, an LRU-based
instantiation could additionally track hit positions in the queue and their reuse distances. These extensions are beyond the scope of this work.
}

\begin{figure}[t]
  \centering
  \begin{subfigure}[t]{0.53\linewidth}
    \centering
    \includegraphics[width=\linewidth]{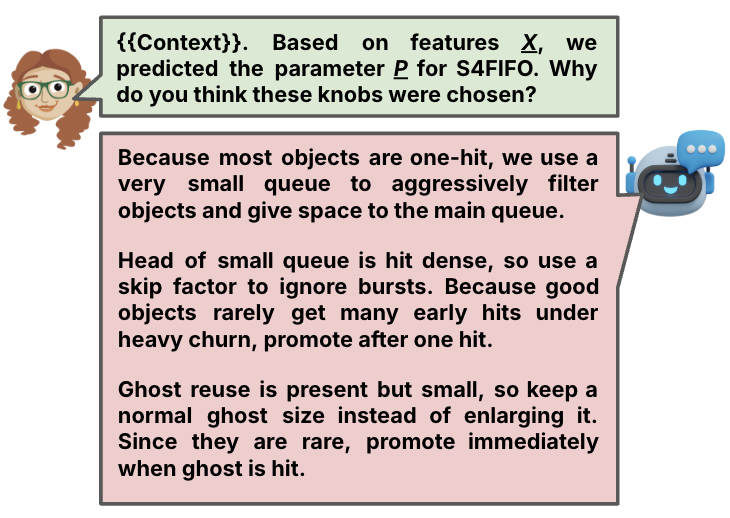}
    \caption{Explaining predicted parameters}
    \label{fig:learning:interpretability}
  \end{subfigure}\hfill
  \begin{subfigure}[t]{0.46\linewidth}
    \centering
    \includegraphics[width=\linewidth]{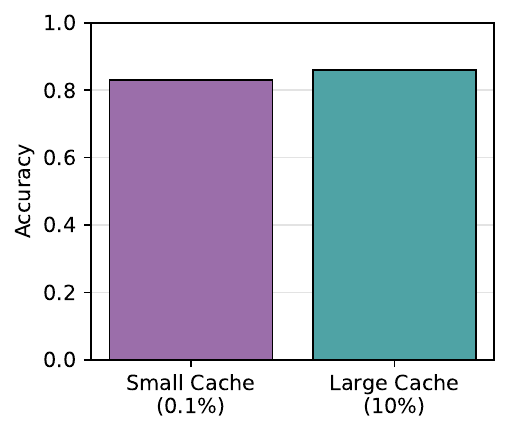}
    \caption{\rev{Selecting suitable parameters}}
    \label{fig:learning:experiment}
  \end{subfigure}
\vspace{-0.5em}
  \caption{
 LLM-based interpretability evaluation.
S4-FIFO's cache-level features and semantic parameters allow an LLM to explain predicted configurations and distinguish suitable parameter choices.}
  \vspace{-1.5em}
  \label{fig:learning:interpretability_eval}
\end{figure}

\vspace{-0.5em}
\subsection{Is the Learned Decision Interpretable? } 
\sysname learns from high-level cache features and predicts high-level knobs. Because both the inputs and outputs operate at the cache level, the decisions it produces are far more interpretable than those of prior smart-cache designs. Unlike approaches that make opaque predictions (e.g., ranking objects or estimating reuse densities), \sysname expresses its decisions using knobs that operators already understand, such as queue sizes and promotion thresholds. The features \sysname used characterize workload, capturing burstiness, one-hit ratios, and reuse-distance patterns, making the mapping from features to knobs easier to interpret. 

\rev{
We use LLM-based evaluation as a proxy for semantic interpretability: if an external reasoning model can connect workload-level features to the selected cache knobs, then the learned decisions are likely expressed in a form that operators can reason about. We evaluate this with two language-model-based experiments, shown in \autoref{fig:learning:interpretability_eval}. First, given workload features and a predicted \sysname configuration, LLMs are able to explain why the selected knobs match the workload behavior. 
Second, we provide an LLM with the workload features and two configurations with distinctly different miss ratios and ask it to identify the better configuration. Across 100 traces for each cache size, the LLM performs substantially above chance, achieving 83\% and 86\% accuracy in the small-cache and large-cache setting. While these results do not establish the correctness of the learned policy, they suggest that \sysname's feature and parameter spaces are semantically meaningful enough to support external reasoning, making the selected configurations easier to explain, justify, and debug. \revv{We provide further details in Appendix \autoref{sec:appendix-interpret}.}
}

\section{Related Work}
\label{sec:related}

\paragraph{Optimizing cache efficiency. }
Many studies have examined the design of eviction algorithms to improve efficiency. 
These include static heuristics such as S3-FIFO~\cite{s3fifo}, SIEVE~\cite{sieve}, QDLP~\cite{yang_fifo_2023}, LIRS~\cite{LIRS, zhong_lirs2_2021}, clock variants~\cite{carr_wsclock_1981, jiang_clock-pro_2005, li_clock-pro_2019}, 2Q~\cite{2Q}, S4LRU~\cite{huang_analysis_2013}, LRFU~\cite{donghee_lee_lrfu_2001}, LFU variants~\cite{arlitt_evaluating_2000, arlitt_performance_2000, karakostas_practical_2000}, greedy-dual variants~\cite{cherkasova_improving_1998, jin_popularity-aware_2000}, MQ~\cite{zhou_multi-queue_2001}, Clock2Q+~\cite{Clock2Q+}, Hyperbolic~\cite{blankstein_hyperbolic_2017}; 
per-miss cache-level adaptive algorithms, such as ARC~\cite{ARC}, CAR~\cite{bansal_car_2004}, SEQ~\cite{glass_adaptive_1997}, EELRU~\cite{smaragdakis_eelru_1999}, Clock-Pro+~\cite{li_clock-pro_2019}, LeCaR~\cite{LeCaR}, CACHEUS~\cite{rodriguez_learning_2021}, TinyLFU~\cite{TinyLFU}; 
and object-level learning algorithms, such as LRB~\cite{LRB}, 3L-Cache~\cite{3L-Cache}, LHD~\cite{LHD}, PA-Cache~\cite{fan_pa-cache_2020}, RL-Belady~\cite{yan_rl-belady_2020}, Raven~\cite{hu_raven_2022}, HALP~\cite{song_halp_2023}, Darwin~\cite{chen_darwin_2023}. Compared with these algorithms, to the best of our knowledge, this is the first work to propose learning-augmented heuristics that separate the control and data-serving planes to achieve both high efficiency and performance. \sysname demonstrates this idea by using a pre-trained model to learn the parameters of S3-FIFO. 



\vspace{-1em}
\paragraph{Machine learning for systems. }
Prior works integrate ML into systems at three distinct levels of the stack, ranging from {replacing internal mechanisms} to {tuning external knobs}. 
(1) Structural replacement: 
Works such as learned indexes~\cite{learnedIndex,learnedIndexPractical} and learned bloom filters~\cite{learnedBloom} aim to \textit{replace} data structures with neural networks. 
(2) Policy optimization: ML replaces specific heuristics (policies) while keeping the underlying mechanism intact. Examples include RL-based flash admission~\cite{cachesack}, and language-model-based C++ memory manager~\cite{llama,llama-ac}. 
(3) Black-box autotuning: Systems like OtterTune~\cite{ottertune} treat the database as a black box, optimizing configuration knobs via Bayesian optimization. 
\sysname occupies a distinct point in this design space: unlike structural replacement, it preserves the mechanism’s deterministic safety (the S4-FIFO data path); unlike per-instance policy optimization, it leverages a pre-trained foundation model to generalize to unseen workloads; and unlike black-box autotuning with slow iterative search, it performs zero-shot inference, quickly selecting an optimal configuration after a brief observation window and avoiding costly online exploration.

\vspace{-1em}
\section{Conclusion}
We introduce Learning-Augmented Heuristics (LAH), which keeps the data path simple while using a pre-trained model to configure a few semantic knobs. 
Our S4-FIFO instantiation extends S3-FIFO with tunable parameters and selects configurations based on cache-level features. 
Evaluated on 1,035 production traces, S4-FIFO achieves higher average miss-ratio reduction and stronger robustness than state-of-the-art algorithms, while matching the throughput of heuristics. These results indicate that learning to configure expressive heuristics is an effective and interpretable way to bring ML into core caching systems. 

\section*{Acknowledgments}
We thank the anonymous reviewers for their valuable feedback and our shepherd for the constructive suggestions. 
We also would like to thank the people and organizations that have open-sourced and shared production traces. 
We thank Cloudlab~\cite{cloudlab} for providing infrastructure support to run experiments.

\bibliographystyle{plain}
\bibliography{reference,jason}

\appendix

\newpage

\section{Appendix}
\subsection{Pre-training Workflow}
\label{appx:appendix-pretrain}
Unlike online learning approaches that train a model from scratch, \sysname utilizes a ``foundation model'' approach. In this section, we introduce how to pre-train the model offline on a massive corpus of diverse traces in detail. 

\paragraph{Trace simulation \& feature collection. } 
We run the simulation with default parameters of S4-FIFO, which aligns with the behavior of S3-FIFO to collect features. Specifically, S4-FIFO warms up with requests until the cache is full. Afterward, the feature collector starts working, continuously monitors the trace, and exports features after a preset number of requests. We set the request number as 20\% of the total requests in the trace. For each epoch, we simulate the cache behavior and record the feature vector $\mathbf{x}_t$ at the decision point. This ensures the model learns to associate runtime signals with future performance.

\paragraph{Oracle labeling} 
To generate the ground truth, we perform an exhaustive brute-force simulation over the entire parameter grid for every trace epoch. 
(1) {Best Configuration ($y^*$):} The configuration that yields the lowest Miss Ratio (MR) for the upcoming epoch is marked as the target label. 
(2) {Cost Matrix Construction:} Crucially, we compute the regret for \textit{every} candidate class relative to the optimal one. This data populates the cost matrix $\mathbf{L}$, enabling the risk-aware loss function described in \autoref{subsec:learningObj}.

\paragraph{Supervised training. } 
We train a GBDT classifier using LightGBM on the collected pairs of $(\mathbf{x}_t, y^*_t)$. The model is optimized to minimize the multi-class Log-Loss, calibrating the output probabilities. By training on a diverse dataset covering block I/O, KV, and CDN workloads, the model learns to identify workload patterns and maps them to a configuration that reduces the miss ratio. 

\paragraph{Request Serving.} After a prediction, we decode the prediction as the new tuned parameters and set it with the configuration interfaces from \sysname. \sysname will automatically handle the change for different queue sizes and promotion thresholds.

\subsection{Model Interpretability}
\label{sec:appendix-interpret}
Given a context prompt and a feature-label pair, summarized explanations given by the language model include:
\begin{itemize}
    \item \textbf{Workload 1.} A very high one-hit ratio and a small cache ratio indicate that
    most objects are not reused and capacity is tight. The model, therefore, selects a 
    \emph{very small} small queue (low $s$) to aggressively filter one-hit objects, and 
    an \emph{easy promotion threshold} (low $m$) so that the rare reusable objects can be 
    promoted quickly before they are evicted.

    \item \textbf{Workload 2.} The hit histograms show strong burstiness (most small-queue hits
    concentrated near the head) and a large effective working set. In this case, the model 
    \emph{expands the small queue} (high $s$) to absorb bursts, \emph{raises the promotion threshold}
    (higher $m$) to prevent burst pollution of the main queue, and \emph{enables skip behavior} 
    (higher $k$) to ignore noisy early hits.

    \item \textbf{Workload 3.} A high fraction of ghost hits and deep ghost histograms indicate 
    significant long-distance reuse. The model responds by \emph{enlarging the ghost queue}
    (higher $g$) to track more evicted objects and by \emph{raising the ghost-promotion threshold}
    (higher $t$) to separate long-range reuse from scans.
\end{itemize}

\subsubsection{Feature-Label Reasoning Prompt}
S3-FIFO is a cache eviction algorithm that, instead of maintaining complex recency or 
frequency structures like LRU, uses FIFO queues as its core mechanism. The cache 
contains a \textit{small queue} that acts as a filter for ``one-hit wonders,'' and a 
\textit{main queue} that stores objects deemed useful for longer-term reuse. Objects 
first enter the small queue; if they are not accessed again shortly, they are evicted 
to free space. If an object receives a second access, it is lazily promoted—only upon 
eviction—into the main queue, where it benefits from longer residency. Frequently 
accessed objects in the main queue are reinserted into the head of the queue. 

The algorithm also maintains a \textit{ghost queue}, which tracks metadata of recently 
evicted objects without storing their contents. If such an object reappears, this 
ghost history allows it to bypass probation and be inserted directly into the main 
queue.

Our work focuses on tuning S3-FIFO through the following parameters:

\begin{enumerate}
    \item \textbf{\texttt{s\_param} (small-queue size ratio).}  
    Controls the proportion of cache devoted to short-term recency.  It may increase up 
    to 100\%. Lower values imply aggressive filtering of new objects, while higher 
    values retain more recently accessed items.

    \item \textbf{\texttt{g\_param} (ghost-queue size).}  
    Specifies the size of the ghost queues relative to total cache capacity.  
    Larger ghost queues track 
    more evicted objects, improving long-range reuse detection. Smaller values reduce 
    overhead but risk missing ghost-based promotion opportunities.

    \item \textbf{\texttt{k\_param} (small-queue skip factor).}  
    Indicates the percentage of accesses in the small queue that are not recorded.  
    Higher values help tolerate bursty workloads; lower values 
    suggest the workload is less bursty.

    \item \textbf{\texttt{m\_param} (promotion threshold from small to main).}  
    Defines how many accesses an object must receive in the small queue before being 
    promoted to the main queue.  Higher values (e.g., 2) make promotion more difficult, allowing 
    the cache to tolerate more two-hit objects before admitting them to the main 
    queue. Lower values promote objects quickly; higher values protect the main 
    queue.

    \item \textbf{\texttt{t\_param} (ghost-to-main promotion threshold).}  
    Sets the required number of ghost hits for an object to be promoted from ghost 
    to main.  
    Setting this to 1 requires one ghost hit 
    before promotion, useful for mitigating cyclic scan patterns. Higher settings 
    are rarely used and are workload-specific.
\end{enumerate}

We also introduce the following workload features:

\begin{itemize}
    \item \textbf{Feature 1: \texttt{ratio}, \texttt{log\_ratio} (cache size ratio).}  
    Indicates the relative size of the cache compared to dataset or system demand.  
    Higher ratios imply ample capacity and easier caching; lower ratios reflect 
    greater memory pressure and stricter eviction requirements.

    \item \textbf{Feature 2: \texttt{H\_s}, \texttt{H\_m}, \texttt{H\_g} (fractional hits).}  
    Show where hits originate: the small queue, main queue, or ghost queue.  
    High \texttt{H\_s} implies bursty or short-term reuse; high \texttt{H\_m} implies 
    stable reuse; high \texttt{H\_g} implies long reuse distance.

    \item \textbf{Feature 3: Raw hit counts.}  
    \texttt{hits\_small}, \texttt{hits\_main}, \texttt{hits\_ghost} capture the absolute 
    distribution of demand across queues.

    \item \textbf{Feature 4: \texttt{hist\_small}, \texttt{hist\_main}, \texttt{hist\_ghost}.}  
    Histogram-based reuse-distance distributions showing where hits occur within 
    queue positions.  
    Front buckets imply short reuse distances; deep buckets indicate long reuse 
    distances.

    \item \textbf{Feature 5: \texttt{rho\_onehit} (one-hit ratio).}  
    Fraction of objects accessed only once. High values reflect streaming or 
    scan-like behavior, suggesting smaller small-queue sizes for rapid eviction.

    \item \textbf{Feature 6: \texttt{rho\_unique} (uniqueness ratio).}  
    Indicates working-set size relative to request stream. High values reflect 
    churn-heavy behavior with weak locality; low values indicate strong reuse.

    \item \textbf{Feature 7: \texttt{total\_hits}, \texttt{total\_misses}, \texttt{total\_reqs}.}  
    Provide global performance and request intensity measurements.
\end{itemize}

Given these features, the task is as follows:

\medskip

\textbf{Given:}
\[
\text{Feature: } X, \qquad \text{Output: } Y
\]

\noindent For each parameter (\texttt{s\_param}, \texttt{g\_param}, \texttt{k\_param}, 
\texttt{m\_param}, \texttt{t\_param}), explain why the given feature(s) would lead to 
selecting that particular parameter setting. Provide:

\begin{enumerate}
    \item A paragraph-length explanation for each parameter, reasoning from features 
    to knob selection.
    \item A short ``key takeaway'' summarizing why the knob was chosen and which 
    feature supports the decision.
\end{enumerate}






\subsubsection{Prompt for Pairwise Configuration-Selection Evaluation}
\label{app:pairwise-config-selection-prompt}

We use the following prompt to evaluate whether a language model can reason about the semantic relationship between workload features and S4-FIFO configurations. The model is given the measured features of a real workload and two candidate configurations. It must select the more appropriate configuration and provide a brief justification.

\small

You are an expert in modern cache-eviction algorithms, specifically the S4-FIFO algorithm.

The S4-FIFO eviction algorithm is built on three physical queues:

\begin{itemize}
\item \textbf{SMALL queue:} Filters one-hit wonders; new objects land here first.
\item \textbf{MAIN queue:} Serves as a long-residency store for objects that prove useful.
\item \textbf{GHOST queue:} Maintains metadata-only records of objects evicted from SMALL. A hit here allows an object to skip probation and go directly to MAIN.
\end{itemize}

S4-FIFO exposes the following tunable knobs:

\begin{itemize}
\item \textbf{\texttt{s\_param} (small-queue size ratio).}
Fraction of the total cache devoted to the SMALL queue. Lower values provide more aggressive filtering of new objects, which is useful for scan-heavy workloads with many one-hit wonders. Higher values retain more recently accessed objects, which is useful when the SMALL queue would otherwise discard reused objects too early.
Grid: \texttt{0.05, 0.1, 0.2, 0.3, 0.5, 0.7, 0.9}.
\item \textbf{\texttt{m\_param} (SMALL-to-MAIN promotion threshold).}  
Required number of hits in SMALL before an evicted object moves to MAIN. Setting \texttt{m\_param = 1} promotes an object that receives a second access and is therefore relatively lenient. Setting \texttt{m\_param = 2} requires one additional confirming hit and protects MAIN from bursty objects.  
Grid: \texttt{1, 2}.

\item \textbf{\texttt{t\_param} (GHOST-to-MAIN promotion threshold).}  
Required number of hits in GHOST before an object skips probation and moves to MAIN. Setting \texttt{t\_param = 0} enables immediate promotion upon the first ghost hit. Setting \texttt{t\_param = 1} delays promotion until a second ghost hit, which can help with cyclic-scan patterns in which objects repeatedly enter and leave the cache without being truly hot.  
Grid: \texttt{0, 1}.

\item \textbf{\texttt{g\_param} (ghost-queue size as a multiple of MAIN).}  
Setting \texttt{g\_param = 0.9} maintains a relatively small ghost queue, which captures short-range reuse at low cost. Setting \texttt{g\_param = 3.0} or \texttt{g\_param = 6.0} maintains a larger ghost queue, which captures long-reuse-distance workloads in which evicted objects reappear much later.  
Grid: \texttt{0.9, 3.0, 6.0}.

\end{itemize}

The workload features are collected during a 20\% warm-up period using the default S4-FIFO configuration. The original 20-bin histogram for each queue has been aggregated into four depth buckets for readability. Each bucket value is the sum of five underlying bins. Therefore, the four buckets for each queue sum to approximately 1, and each value directly represents the share of hits falling within that depth range:

\begin{itemize}
\item \texttt{hist\_X\_0-5}: share of hits in bins 0--4, representing the front, head, or most-recent portion of the queue.
\item \texttt{hist\_X\_5-10}: share of hits in bins 5--9.
\item \texttt{hist\_X\_10-15}: share of hits in bins 10--14.
\item \texttt{hist\_X\_15-20}: share of hits in bins 15--19, representing the tail or about-to-be-evicted portion of the queue.
\end{itemize}

Here, \texttt{X} is one of \texttt{small}, \texttt{main}, or \texttt{ghost}. A front-heavy histogram indicates short reuse distances, while a tail-heavy histogram indicates long reuse distances or queue churn. The ghost histogram is particularly useful: hits in deeper ghost buckets indicate that evicted objects return after longer intervals and may therefore justify a larger ghost queue.

The histogram features are presented depth-first across queues: SMALL, MAIN, and GHOST at depth 0--5, followed by SMALL, MAIN, and GHOST at depth 5--10, and so on.

The scalar features are:

\begin{itemize}
\item \texttt{log\_C}: logarithm of the cache capacity, providing a scale signal.
\item \texttt{ratio}: cache size divided by working-set size.
\item \texttt{H\_s}, \texttt{H\_m}, and \texttt{H\_g}: fractions of all hits served by SMALL, MAIN, and GHOST, respectively.
\item \texttt{rho\_onehit}: one-hit-wonder ratio. A high value indicates scan-like behavior.
\item \texttt{rho\_unique}: unique-object ratio. A high value indicates churn or weak locality.
\item \texttt{hits\_small}, \texttt{hits\_main}, and \texttt{hits\_ghost}: raw hit counts for the three queues.
\item \texttt{total\_hits}, \texttt{total\_misses}, and \texttt{total\_reqs}: global counters.
\end{itemize}

If MAIN already serves most hits, as indicated by a high \texttt{H\_m} value and many MAIN hits across depths, expanding SMALL is risky. A larger SMALL queue reduces the MAIN budget and may evict useful MAIN objects. In this regime, prefer keeping SMALL modest.

When asked to choose between two options labeled A and B, note that the labels are assigned randomly: A is no more likely to be correct than B. Evaluate both options on their merits. Do not equivocate.

Given \textbf{Workload features} X, and \textbf{Two candidate S4-FIFO configurations: A and B}, which configuration is more appropriate for this workload?

Answer A or B.

\subsection{Artifact Appendix}


This artifact contains the implementation and evaluation infrastructure for
S4-FIFO. The artifact includes a \texttt{libCacheSim}-based implementation for
miss-ratio evaluation, a Meta CacheLib integration for throughput benchmarking,
model-training and analysis scripts, a Docker-based environment, and
documentation for reproducing the experiments.

\paragraph{Scope.}
The artifact can be used to validate the following claims:

\begin{itemize}
\item \textbf{Miss-ratio evaluation.}
The \texttt{libCacheSim}-based simulator can compare S4-FIFO against
S3-FIFO and other cache-eviction algorithms, including FIFO, LRU, 2Q, ARC,
LeCaR, LIRS, SIEVE, and LFU, on supported traces.

\item \textbf{Learning pipeline.}
The artifact includes scripts for parameter search, feature collection,
representative-configuration selection, model training, model evaluation,
feature-importance analysis, and cross-dataset analysis.

\item \textbf{Throughput evaluation.}
The CacheLib integration can be used to evaluate the throughput overhead of
S4-FIFO and compare it against optimized heuristic implementations using
CacheBench.

\item \textbf{Extensibility.}
The artifact can be used to evaluate custom traces, inspect the predicted
configuration, and manually modify S4-FIFO parameters.

\end{itemize}

The paper evaluates S4-FIFO using 5,175 production traces from 14 sources. Some
of these traces contain proprietary data and are not distributed publicly.

\paragraph{Contents.}

The repository contains the following main components:

\begin{itemize}
\item \texttt{libCacheSim/}: simulator-based implementation of S4-FIFO and
baseline algorithms;
\item \texttt{CacheLib/}: CacheLib integration for throughput evaluation;
\item \texttt{analysis/}: scripts for parameter search, feature analysis,
model training, and result analysis;
\item \texttt{doc/}: detailed reproduction instructions;
\item \texttt{Dockerfile}: containerized environment; and
\item \texttt{Makefile}: convenience commands for building and testing.
\end{itemize}

\paragraph{Hosting.}

The artifact is publicly hosted on GitHub:

\begin{center}
\url{https://github.com/cacheMon/osdi26-s4-fifo}
\end{center}

The artifact version corresponding to this paper is:

\begin{itemize}
\item Branch: \texttt{osdi26}
\item Commit: \texttt{000095fc3c96a95ff1b9a19d101fb1411c7f4d54}
\end{itemize}


\end{document}